\documentclass[twocolumn]{aastex631}

\usepackage{epsfig}
\usepackage{graphicx}
\usepackage{amsthm}
\usepackage{amssymb}
\usepackage{rotating}
\usepackage{multirow}
\usepackage{longtable}
\usepackage{savesym}
\savesymbol{tablenum}
\restoresymbol{SIX}{tablenum}
\usepackage{booktabs}
\usepackage{mathtools}

\defcitealias{2019ApJ...873...51L}{Paper~I}
\defcitealias{2020ApJ...888...98L}{Paper~II}
\defcitealias{2021DeBuizer}{Paper~III}
\defcitealias{2022DeBuizer}{Paper~IV}

\accepted{March 22, 2023 in ApJ}

\shorttitle{Surveying G\ion{H}{2} Regions: V. DR7 and K3-50}
\shortauthors{De Buizer et al.}

\begin{document}

\title{Surveying the Giant \ion{H}{2} Regions of the Milky Way with SOFIA: V. DR7 and K3-50}

\correspondingauthor{James De Buizer}
\email{jdebuizer@sofia.usra.edu}

\author[0000-0001-7378-4430]{James M. De Buizer}
\affil{SOFIA-USRA, NASA Ames Research Center, MS 232-12, Moffett Field, CA 94035, USA}

\author[0000-0003-4243-6809]{Wanggi Lim}
\affil{SOFIA-USRA, NASA Ames Research Center, MS 232-12, Moffett Field, CA 94035, USA}

\author[0000-0003-0740-2259]{James T. Radomski}
\affil{SOFIA-USRA, NASA Ames Research Center, MS 232-12, Moffett Field, CA 94035, USA}

\author[0000-0001-6159-2394]{Mengyao Liu}
\affil{Dept. of Astronomy, University of Virginia, Charlottesville, Virginia 22904, USA}

\begin{abstract}
We present our fifth set of results from our mid-infrared imaging survey of Milky Way Giant \ion{H}{2} (G\ion{H}{2}) regions with our detailed analysis of DR7 and K3-50. We obtained 20/25 and 37\,$\mu$m imaging maps of both regions using the FORCAST instrument on the Stratospheric Observatory For Infrared Astronomy (SOFIA). We investigate the multi-scale properties of DR7 and K3-50 using our data in conjunction with previous multi-wavelength observations. Near to far-infrared spectral energy distributions of individual compact infrared sources were constructed and fitted with massive young stellar object (MYSO) models. We find eight out of the ten (80\%) compact sources in K3-50 and three out of the four (75\%) sources in DR7 are likely to be MYSOs. We derived luminosity-to-mass ratios of the extended radio sub-regions of DR7 and K3-50 to estimate their relative ages. The large spread in evolutionary state for the sub-regions in K3-50 likely indicates that the star-forming complex has undergone multiple star forming events separated more widely in time, whereas the smaller spread in DR7 likely indicates the star formation sub-regions are more co-eval. DR7 and K3-50 have Lyman continuum photon rates just above the formal threshold criterion for being categorized as a G\ion{H}{2} region (10$^{50}$ photons/s) but with large enough errors that this classification is uncertain. By measuring other observational characteristics in the infrared, we find that K3-50 has properties more akin to previous bonafide G\ion{H}{2} regions we have studied, whereas DR7 has values more like those of the non-G\ion{H}{2} regions we have previously studied. 
\end{abstract}

\keywords{ISM: \ion{H}{2} regions --- infrared: stars —-- stars: formation —-- infrared: ISM: continuum —-- ISM: individual(DR7, W58A, K3-50, Sh2-100)}

\section{Introduction} 

This is the fifth paper in a series of studies of the infrared properties of galactic giant \ion{H}{2} (G\ion{H}{2}) regions using 20/25 and 37\,$\mu$m mid-infrared imaging data from the Stratospheric Observatory for Infrared Astronomy (SOFIA). G\ion{H}{2} regions represent the most luminous ionized regions in our Galaxy, and are defined as possessing a Lyman continuum photon rate of greater than $N_{LyC} = 10^{50}$\,photons/s \citep{2004MNRAS.355..899C, 1970IAUS...38..107M}. These regions are believed to be where the largest OB star clusters in the Galaxy have formed and are presently continuing to form. In our first three papers, we performed an in-depth study of three of the top six most luminous (in terms of $N_{LyC}$) G\ion{H}{2} regions in the Galaxy, G49.5-0.4 in W51A\footnote{W51A is comprised of two G\ion{H}{2} regions, the highly luminous G49.50-0.4 and the $\sim$6$\times$ less luminous G49.4-0.3, both of which are studied in \citetalias{2019ApJ...873...51L}.} (Lim \& De Buizer 2019; hereafter ``\citetalias{2019ApJ...873...51L}''), M17 (Lim et al. 2020; hereafter ``\citetalias{2020ApJ...888...98L}''), and W49A (De Buizer et al. 2021; hereafter ``\citetalias{2021DeBuizer}''). We followed up those papers with a study that reassessed the Milky Way G\ion{H}{2} census of \citet{2004MNRAS.355..899C} using the most recent data on the distances to each source, which culled the list from 56 to 42 G\ion{H}{2} candidates (De Buizer et al. 2022; hereafter ``\citetalias{2022DeBuizer}''). In that paper we discussed two sources for which we had obtained SOFIA data, Sgr\,D and W42, but for which our reassessment concluded were not powerful enough to be considered G\ion{H}{2} regions given their lower Lyman continuum photon rate calculated at their updated distances. It was also shown in \citetalias{2022DeBuizer} that Sgr\,D and W42 possessed observational and physical characteristics different from the more luminous G\ion{H}{2} regions previously studied as a part of this survey (i.e., W51A:G49.5-0.4, M17, and W49A). Perhaps the most striking difference is that both Sgr\,D and W42 appear to be fully or dominantly ionized and/or heated by a single massive star.

\citetalias{2022DeBuizer} also determined that 11 of the 42 G\ion{H}{2} regions in the census had values above the $N_{LyC} = 10^{50}$\,photons/s criterion, however their lower limit errors do go below the cut-off value. Such sources were categorized as ``likely'' G\ion{H}{2} regions. In this present paper we will concentrate on two of these G\ion{H}{2} region ``edge cases'' for which we have obtained SOFIA data: DR7 with log$N_{LyC} = 50.10^{+0.14}_{-0.18}$\,photons/s, and K3-50 with log$N_{LyC} = 50.07^{+0.19}_{-0.17}$\,photons/s. We will compare their physical properties and morphological characteristics to the previously studied highly luminous G\ion{H}{2} regions studied in our first three papers in this series, as well as compare them to the sources which fall below the G\ion{H}{2} region threshold that were studied in \citetalias{2022DeBuizer}.

DR7 is a lesser known source seen in projection to lie within the confines of the well-studied Cygnus X region, however it is not believed to be a member of the main Cygnus X complex, and is instead thought to lie in the Perseus Arm behind it \citep[e.g.,][]{1988A&A...191..313P, 2011ApJ...738...27B}. Given its brightness and large distance, DR7 is about 4 times more powerful than Orion A, but this is still only about a twentieth as powerful as W49A \citep{1991A&A...241..551W}. The second source covered in this paper is K3-50, which is the brightest sub-region of the much larger W58 giant molecular cloud complex, and it contains at its heart a very well-known compact \ion{H}{2} region named K3-50A which has been the subject of many studies \citep[see][and references therein]{2010ApJ...714.1015S}.  

DR7 is morphologically very different in appearance from K3-50. The latter is a collection of several radio emitting regions that together combine to create the overall G\ion{H}{2}-level of emission, making it analogous to the subjects of \citetalias{2019ApJ...873...51L}, G49.5-0.4 and G49.4-0.3 in W51A. In contrast, DR7 is a large, almost contiguous ionized region tracing the edge of a giant cavity, much in the same manner as the subject of \citetalias{2020ApJ...888...98L}, M17.

In the next section (Section \ref{sec:obs}), we will discuss the new SOFIA observations and give information on the data obtained for DR7 and K3-50. In Section \ref{sec:results1}, we will give more background on these G\ion{H}{2} regions as we compare our new data to previous observations and discuss individual sources and sub-regions within these G\ion{H}{2} regions in-depth. In Section \ref{sec:data}, we will discuss our data analysis, modeling, and derivation of physical parameters of sources and sub-regions. We will compare and contrast the properties of the two G\ion{H}{2} regions and compare those results to the results from the G\ion{H}{2} regions that were the subjects of our previous papers. Our conclusions are summarized in Section \ref{sec:sum}.

\section{Observations and Data Reduction} \label{sec:obs}

The observational techniques and data reduction processes employed on the data were, for the most part, identical to those described in \citetalias{2019ApJ...873...51L} for W\,51\,A. We will briefly detail below some of the salient information regarding how the observations were obtained and highlight the reduction details specific to these new observations. For a more in-depth discussion of the details and techniques employed, refer to \citetalias{2019ApJ...873...51L}.

FORCAST is a dual-array mid-infrared camera capable of taking simultaneous images at two wavelengths. The short wavelength camera (SWC) is a 256$\times$256 pixel Si:As array optimized for $5-25\,\mu$m observations; the long wavelength camera (LWC) is a 256$\times$256 pixel Si:Sb array optimized for $25-40\,\mu$m observations. After correction for focal plane distortion, FORCAST effectively samples at 0$\farcs$768/pixel, which yields a 3$\farcm$04$\times$3$\farcm$02 instantaneous field of view. 

Data were taken for DR7 on the night of November 4, 2015 (SOFIA Cycle 3, Program ID 03\_0008) on Flight 254 at a flight altitude of 41000 ft. Observations were obtained using the 20\,$\mu$m ($\lambda_{eff}$=19.7\,$\mu$m; $\Delta\lambda$=5.5\,$\mu$m) and 37\,$\mu$m ($\lambda_{eff}$=37.1\,$\mu$m; $\Delta\lambda$=3.3\,$\mu$m) filters simultaneously using an internal dichroic, with an on-source exposure time of 295\,s. The bright mid-infrared-emitting region of DR7 is larger than a single FORCAST field, and thus required two pointings which were mosaicked together to cover the whole source. Images from each individual pointing were stitched together using the SOFIA Data Pipeline software REDUX \citep{2015ASPC..495..355C} into a final mosaic (a ``Level 4'' SOFIA data product).

K3-50 was first observed on July 4, 2014 (SOFIA Cycle 2, Program ID 02\_0113) on Flight 176 at an altitude of 39000 ft. The presence of clouds seen in the data and weather in the area forced us to stop early and close the telescope cavity door for the remainder of the planned observing time. We were only able to obtain 120\,s of total on-source exposure time on the field that only covered the K3-50 C region. Unlike the DR7 observations, these were taken with the 25\,$\mu$m ($\lambda_{eff}$=25.3\,$\mu$m; $\Delta\lambda$=1.9\,$\mu$m) filter in the short wavelength camera, but still employed the same 37\,$\mu$m filter in the long wavelength camera. As we will discuss later, the flux calibration of these data are less certain.

We revisited K3-50 on September 16, 2015 (SOFIA Cycle 3, Program ID 03\_0008) on Flight 239 at a flight altitude of 41000 ft. We were able this time to obtain images of a field centered on K3-50 A through the 20 and 37\,$\mu$m filters with an 334\,s exposure time. This field did not cover the K3-50 C region. The C region was to be re-observed on a later flight, but that flight was canceled. 

Flux calibration for the DR7 and K3-50 data was provided by the SOFIA Data Cycle System (DCS) pipeline and the final total photometric errors in the images were derived using the same process described in \citetalias{2019ApJ...873...51L}. Except for the K3-50 C field, the estimated total photometric errors are 15\% for 20\,$\mu$m and 10\% for 37\,$\mu$m. 

All images then had their astrometry absolutely calibrated using Spitzer data by matching up the centroids of point sources in common between the Spitzer and SOFIA data. Absolute astrometry of the final SOFIA images is assumed to be better than 1$\farcs$5, which is a slightly more conservative estimate than that quoted in \citetalias{2019ApJ...873...51L} (i.e., 1$\farcs$0) due to slight changes in the focal plane distortion and our ability to accurately correct it with the limited calibration data available for these observations.

For the K3-50 C region data, where data collection was halted due to clouds, we looked at the individual 40s frames that constituted the final 120\,s coadded image. We also looked at the four fits extensions to the images that contain the raw chop data which one can use to yield a determination of the background sky emission level during the observation. We determined that only the first (40s on-source time) frame appeared to have stable sky emission levels consistent with nominal (i.e., photometric condition) data at both 25 and 37\,$\mu$m (as they were taken simultaneously with a dichroic), while the other three 40s frames had high and variable sky emission levels indicative of clouds. Since sources K3-50 A and B were also on the field, we used the 37\,$\mu$m photometry of these two sources from data obtained in Cycle 3 as a comparison and found that the A and B photometry values for the good 40s frame were depressed by 15\% for both sources A and B. We therefore adjusted the photometric values of the C sources that we obtained from the good 40s image by +15\% and believe that this ``bootstrap" calibration is as accurate as the nominal 37\,$\mu$m data (i.e., 10\% uncertainty). For the 25\,$\mu$m data, we had no Cycle 3 25\,$\mu$m data (only 20\,$\mu$m) and so a similar bootstrap method of photometric correction from Cycle 3 data could not be performed. However, clouds tend to affect/depress the longer wavelength flux data more than the shorter wavelength data, so it is assumed that the photometry of the first 40s 25\,$\mu$m frame is likely less affected than the 37\,$\mu$m data. Assuming that source fluxes at 20 and 25\,$\mu$m will be similar but not exactly the same (their filter transmission profiles do overlap modestly), as a sanity check we compared the photometry of sources A and B on the good 40s image at 25\,$\mu$m to that of the Cycle 3 20\,$\mu$m data and found that they agree to within 10\%. We therefore believe that the 25\,$\mu$m photometry of the K3-50 C region (which encompasses sources C1 and C2) is reliable to within 25\%.  

\section{Comparing SOFIA Images to Previous Imaging Observations} \label{sec:results1}

\subsection{DR7}\label{sec:DR7}

\begin{figure*}[tb!]
\epsscale{1.15}
\plotone{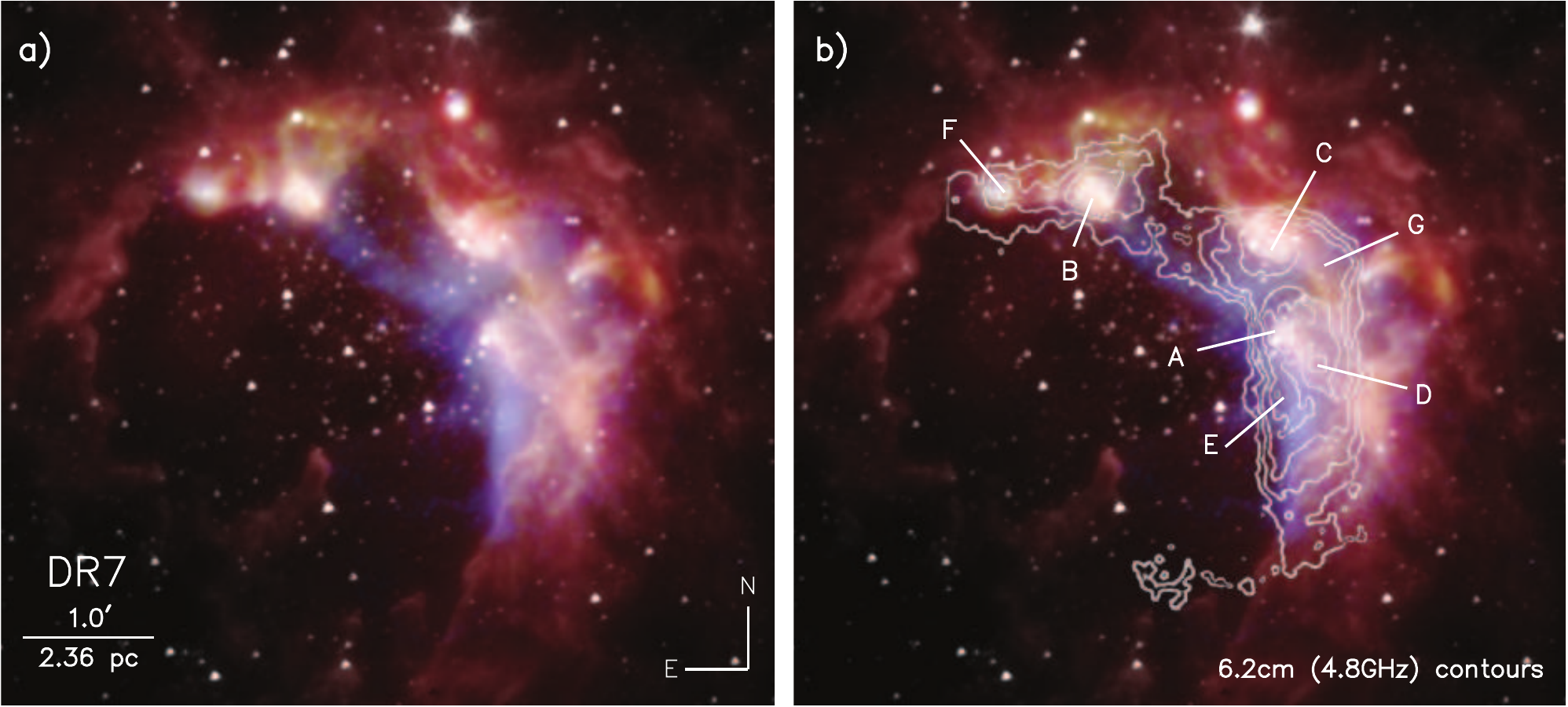}
\caption{Images of DR7. a) A 3-color image of a $\sim6\farcm0\times 5\farcm$0 field centered on DR7 where blue is the SOFIA-FORCAST 20\,$\mu$m image, green is the SOFIA-FORCAST 37\,$\mu$m image, and red is the Herschel-PACS 70\,$\mu$m image. Overlaid in white is the Spitzer-IRAC 3.6\,$\mu$m image, which traces the revealed stars within DR7, field stars, and hot dust. A white horizontal bar in the lower left corner shows the scale of the image. b) The same 3-color image of DR7 with the 6.2\,cm (4.8\,GHz) radio continuum contours of \citet{1986ApJ...306..122O} overlaid in white with the seven previously identified radio sub-regions labeled.\label{fig:fig1}}
\end{figure*}

The name, DR7, comes from \citet{1966ApJ...144..937D} who mapped the Cygnus X region at 6\,cm with $\sim$10$\arcmin$ resolution, though this source had previously been resolved at 21\,cm by \citet{1964ApJ...139..545P}. DR7 was further resolved into seven radio continuum peaks by \citet{1974IAUS...60..219B} using Very Large Array (VLA) data at 6\,cm, and later confirmed by \citet{1986ApJ...306..122O}, who labeled those peaks A through G.

Radial velocity measurements of the H100$\alpha$ recombination line in the \ion{H}{2} region of DR7 by \citet{1988A&A...191..313P} discovered that the gas there has a distinctly different local standard of rest velocity ($v_{lsr} \sim -41$\,km/s) than the rest of the gas in the Cygnus X region ($-12 < v_{lsr} < +10$\,km/s). They calculated that DR7 is likely to exist behind the Cygnus X region, most likely placing it in the more distant Perseus Arm. Indeed, \citet{1993ApJ...405..706O} show that there are three arms in projection in the Cygnus X region; the Local Arm with sources at $<$5\,kpc, the Perseus Arm from $7-9.5$\,kpc, and the Outer Arm, from $10-12$\,kpc. Nonetheless, there does seem to be confusion in the literature with several papers after these results claiming that DR7 is as close as 1.5\,kpc \citep{1994ApJS...91..659K}, which would place it at a distance consistent with that derived for the Cyg OB 2 cluster \citep[1.7\,kpc;][] {2000A&A...360..539K}. Though they measure the H110$\alpha$ recombination line at $-41$\,km/s, \citet{2011A&A...532A.127D} assign DR7 to the tangent position at 1.56\,kpc. However, \citet{2011ApJ...727..114R} claim that it is unlikely that DR7 is a source with an extremely high peculiar velocity but located at the same distance as the rest of the Cygnus X region due to it being the only region that lacks a signature in the extinction maps (created from 2MASS stars) by \cite{2007A&A...476.1243M}. Here we adopt the value from our work in \citetalias{2022DeBuizer}, where we calculated the kinematic distance to DR7 to be $7.30^{+0.84}_{-0.72}$\,kpc based upon the highest precision line observations we could find, namely the H91$\alpha$ transition measurements ($v_{lsr} = -39.17\pm0.07$\,km/s) of \citet{2006ApJ...653.1226Q}. 

Given their revelation that DR7 is likely more distant than the bulk of the Cygnus X region, \citet{1988A&A...191..313P} were the first to also realize the main ramification of that distance change; namely, that DR7 is not a modestly bright radio continuum region, but instead perhaps a more powerful giant \ion{H}{2} region. In fact, in \citetalias{2022DeBuizer} we show that at our adopted distance DR7 must have a Lyman continuum photon rate of log$N_{LyC}=50.10^{+0.14}_{-0.18}$\,photons/s to account for its radio flux. 

\citet{1994ApJS...91..659K} were the first to make sub-arcsecond radio continuum images of DR7 with the VLA, taking data which had an image resolution of 0$\farcs$5 at 2\,cm and 0$\farcs$9 at 3.6\,cm. \citet{1994ApJS...91..659K} identified two compact sources in the region. The first, named G79.321+1.291, is considered to be an irregular compact \ion{H}{2} region with an integrated 3.6\,cm flux density of 6.4\,mJy. This source is coincident with the western part of the elongated source B from \citet{1986ApJ...306..122O} seen at 6.2\,cm. The second source, G79.320+1.313, is seen as an unresolved source with a flux density of 6.5\,mJy at 2.0\,cm and 3.9\,mJy at 3.6\,cm, and is not in the observed field of \citet{1986ApJ...306..122O}, though there is a source seen here in the 6.2\,cm maps of \citet{1974IAUS...60..219B}, but is not labeled or referenced. 

\begin{figure*}[tb!]
\epsscale{1.15}
\plotone{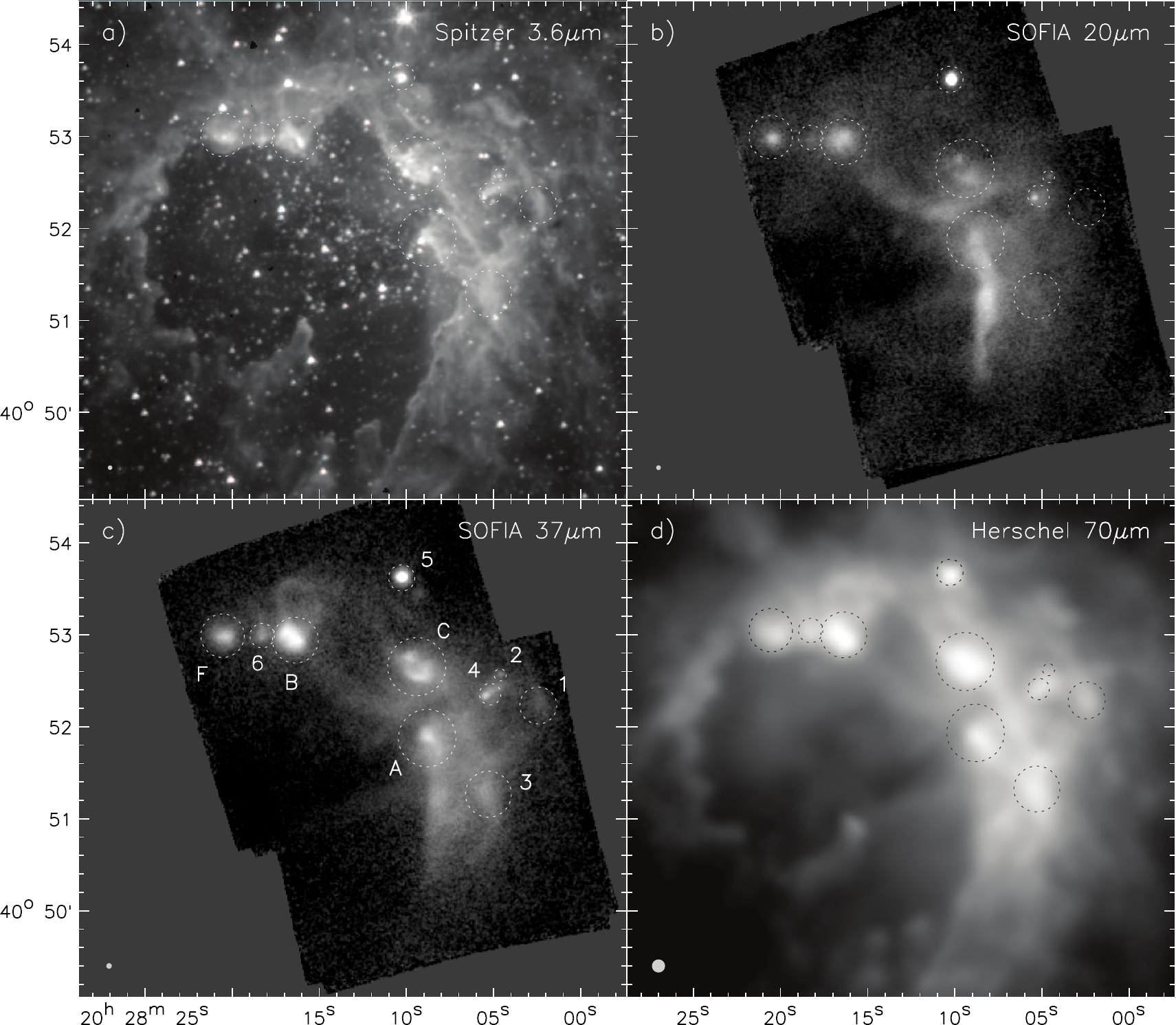}
\caption{Images of DR7 at a) Spitzer-IRAC 3.6\,$\mu$m, b) SOFIA-FORCAST 20\,$\mu$m, c) SOFIA-FORCAST 37\,$\mu$m, and d) Herschel-PACS 70\,$\mu$m. The new infrared sources are numbered and the infrared sources that are previously identified radio sub-regions are lettered. The sizes and locations of the circles around the sources corresponds to the coordinates and apertures used for photometry at 37\,$\mu$m (Table \ref{tb:DR7_sources}). The gray dot in the lower left corner of each panel shows the spatial resolution at the given wavelength. \label{fig:fig2}}
\end{figure*}

\begin{deluxetable*}{rccclllllc}
\tabletypesize{\scriptsize}
\tablecolumns{10}
\tablewidth{0pt}
\tablecaption{SOFIA-FORCAST Observational Parameters of Infrared Sources in DR7}
\tablehead{\colhead{ }&
           \colhead{  } &
           \colhead{  } &
           \multicolumn{3}{c}{${\rm 20\mu{m}}$}&
           \multicolumn{3}{c}{${\rm 37\mu{m}}$} &
           \colhead{}\\
           \cmidrule(lr){4-6} \cmidrule(lr){7-9}\\
           \colhead{ Source }&
           \colhead{ R.A.(J2000) } &
           \colhead{ Dec.(J2000) } &
           \colhead{ $R_{\rm int}$ } &
           \colhead{ $F_{\rm int}$ } &
           \colhead{ $F_{\rm int-bg}$ } &
           \colhead{ $R_{\rm int}$ } &
           \colhead{ $F_{\rm int}$ } &
           \colhead{ $F_{\rm int-bg}$ } &
           \colhead{Aliases}\\
	   \colhead{  } &
	   \colhead{  } &
	   \colhead{  } &
	   \colhead{ ($\arcsec$) } &
	   \colhead{ (Jy) } &
	  \colhead{ (Jy) } &
	   \colhead{ ($\arcsec$) } &
	   \colhead{ (Jy) } &
	   \colhead{ (Jy) } &
	   \colhead{}  
}
\startdata
 \multicolumn{10}{c}{Compact Sources} \\
 \hline
 DR7 2 & 20:28:04.6 & +40:52:34.0 &4	    &0.91	        &0.37	 &4	    &2.57	&1.23  &   \\
 DR7 4 & 20:28:05.1 & +40:52:21.4 &7	    &5.80	        &3.22	 &7	    &18.6	&11.5  &   \\
 DR7 5 & 20:28:10.2 & +40:53:37.0 &8	    &16.6	        &15.9	 &9	    &55.1	&49.8  & G79.320+1.313  \\ 
 DR7 6 & 20:28:18.3 & +40:52:58.4 &8	    &3.37	        &0.90	 &8	    &19.9	&8.13  &   \\  
 \hline
 \multicolumn{10}{c}{Extended Sources} \\
 \hline
 DR7 1 & 20:28:02.3 & +40:52:14.8 &12	&$<$0.44	    &\nodata &12	&6.44	&6.16  &   \\
 DR7 3 & 20:28:05.2 & +40:51:17.3 &15	&$<$14.9	    &\nodata &15	&64.8	&14.3  &   \\ 
 DR7 A     & 20:28:08.8 & +40:51:53.9 &19	&67.5	        &28.8    &19	&174	&147  &   \\ 
 DR7 B     & 20:28:16.5 & +40:52:57.3 &15	&25.8	        &19.6	 &15	&162	&131  & G79.321+1.291  \\ 
 DR7 C     & 20:28:09.4 & +40:52:38.5 &19	&37.1	        &20.2	 &19	&185	&123  &   \\ 
 DR7 F     & 20:28:20.5 & +40:52:58.9 &12	&9.90	        &6.73	 &15	&56.4	&41.9  &    
\enddata
\tablecomments{\footnotesize R.A. and Dec. are for the center of the photometric apertures used. $R_{\rm int}$ indicates radius of the aperture. $F_{\rm int}$ is the integrated flux in the aperture, while $F_{\rm int-bg}$ is that same flux with an estimate of the surrounding background subtracted. See Section \ref{sec:data} for more information.}
\label{tb:DR7_sources}
\end{deluxetable*}

Of the larger radio sources identified by \citet{1974IAUS...60..219B} and \citet{1986ApJ...306..122O}, we see clear 20 and 37\,$\mu$m infrared dust continuum associated with A, B, C, and F (Figure\,\ref{fig:fig1}). We see a ridge of infrared emission at the location of E, but there is no definitive peak. Sources D and G do not correspond to any infrared emission peaks, and, to the contrary, appear to be where there is a decrease in infrared emission. As for the compact radio continuum sources seen by \citet{1994ApJS...91..659K}, we detect emission at the location of G79.321+1.291, since it is part of larger B region. The brightest source on the SOFIA field at 20 and 37\,$\mu$m is the point-like source associated with G79.320+1.313, which we label as source 5 in Figure\,\ref{fig:fig2}.  Overall, the extended emission in the SOFIA data is arc-shaped, with an apex to the northwest. 

\citet{2002A&A...392..869L} used 2MASS data to find a revealed near-infrared cluster of stars (which they name Cl09) whose center is interior to the arc of infrared emission. If this cluster is indeed at the distance of the Perseus Arm, then they find 44$\pm$15 OB stars, with 8$\pm$5 O stars, and a cluster mass in the range of $1910-4620$\,M$_{\sun}$. The Spitzer-IRAC images of DR7 show the same bright arc seen in radio and the SOFIA data, however, they also show that there is fainter infrared emission completely surrounding the Cl09 star cluster, leading to a bubble-like appearance (i.e., Figure\,\ref{fig:fig2}a). It is therefore likely that the arc we see in the radio and mid-infrared is only the brightest part of a bubble that has been blown out in all directions by the more evolved Cl09 cluster. It is unclear why the northwest side would be so much brighter in the infrared, but it could be that this side is running into denser cloud material causing it to collapse and trigger a new generation of star formation which is occurring in the brighter, dense knots seen in the radio and infrared and helping to heat this part of the bubble. 

\subsection{K3-50 (a.k.a. W58A, Sh2-100)}\label{sec:K350}

W58 is an expansive \citep[1.3$\arcdeg\times$1.6$\arcdeg$;][]{1972A&AS....5..369F} and strong galactic radio region first discovered by \citet{1958BAN....14..215W}. The strongest radio continuum emission lies in the vicinity of the optical emission nebula NGC 6857, and this sub-region of W58 was further resolved into two radio continuum regions by \citet{1959ApJS....4..257S} named  Sh2-99 and Sh2-100, with the latter being the brighter source. Within Sh2-100, \citet{1965BAICz..16..221K} identified a new optical component $\sim$1$\arcmin$ north of NGC 6857 coincident with the peak seen in radio continuum emission and misclassified it as a planetary nebula, leading to the region's moniker of K3-50 (it was source number 50 in Table 3 of their study). The radio emission centered on the position of K3-50 was then further resolved into a group of four smaller \ion{H}{2} regions named A through D, with D being the radio continuum component associated with NGC 6857 and A being the brightest radio peak associated with the location of the misclassified planetary nebula.  Later, \citet{1975MNRAS.170..139H} further resolved C into two separate sources, C1 and C2 and \citet{1976A&A....48..193I} found two new, but far less prominent, radio continuum sources nearby, one about two arcminutes northwest of C named K3-50 E and another two arcminutes west of that, named K3-50 F. \citet{2004MNRAS.355..899C} identify the combined emission from K3-50 A-D as a G\ion{H}{2} in their census, and refer to it as W58A. The region K3-50 A is very prominent in the mid-infrared to radio and has been the subject of numerous studies, but the other regions have been explored far less. For an excellent paper reviewing previous studies all of these K3-50 regions as well as presenting multi-wavelength analyses, we refer the reader to \citet{2010ApJ...714.1015S}. 

The distance to K3-50 has been kinematically derived by multiple studies, with almost all derived values falling in the range of 7.3 to 9.3\,kpc (e.g., \citealt{1975MNRAS.170..139H}; \citealt{2011ApJ...736..149G}; \citealt{2011ApJ...738...27B}; also see discussion in \citealt{2015MNRAS.453.2622B}), with the exception of \citet{2011A&A...532A.127D} who claim the region lies at a kinematic tangent point 2.83\,kpc away. \citet{2010ApJ...714.1015S} were able to spectrally classify the star responsible for ionizing the K3-50 D \ion{H}{2} region (as an O4V star) and based upon its brightness and estimated extinction a distance of $\sim$8.5\,kpc was derived. \citet{2010ApJ...714.1015S} caution that such estimates are highly dependent upon the assumed $M_V$ values of a typical O4V star, which can lead to distances anywhere from 7.9 to 10\,kpc. In \citetalias{2022DeBuizer} we derive a kinematic distance of $7.64^{+0.81}_{-0.54}$\,kpc to this region based upon the highest precision line measurements available (i.e., the H91$\alpha$ measurements of \citealt{2006ApJ...653.1226Q} which have $v_{lsr} = -23.11\pm0.08$\,km/s). This distance estimate is quoted with greater precision than (but still within the combined errors of) those of \citet{2010ApJ...714.1015S}, so we will adopt the 7.64\,kpc value in this work. Note that at the distance of 8.5\,kpc (and its associated errors) estimated by \citet{2010ApJ...714.1015S} the case for K3-50 being a G\ion{H}{2} region would be slightly more robust (i.e., log$N_{LyC} = 50.21^{+0.13}_{-0.22}$\,photons/s, instead of log$N_{LyC} = 50.07^{+0.19}_{-0.17}$\,photons/s).

\begin{figure*}[tb!]
\epsscale{1.16}
\plotone{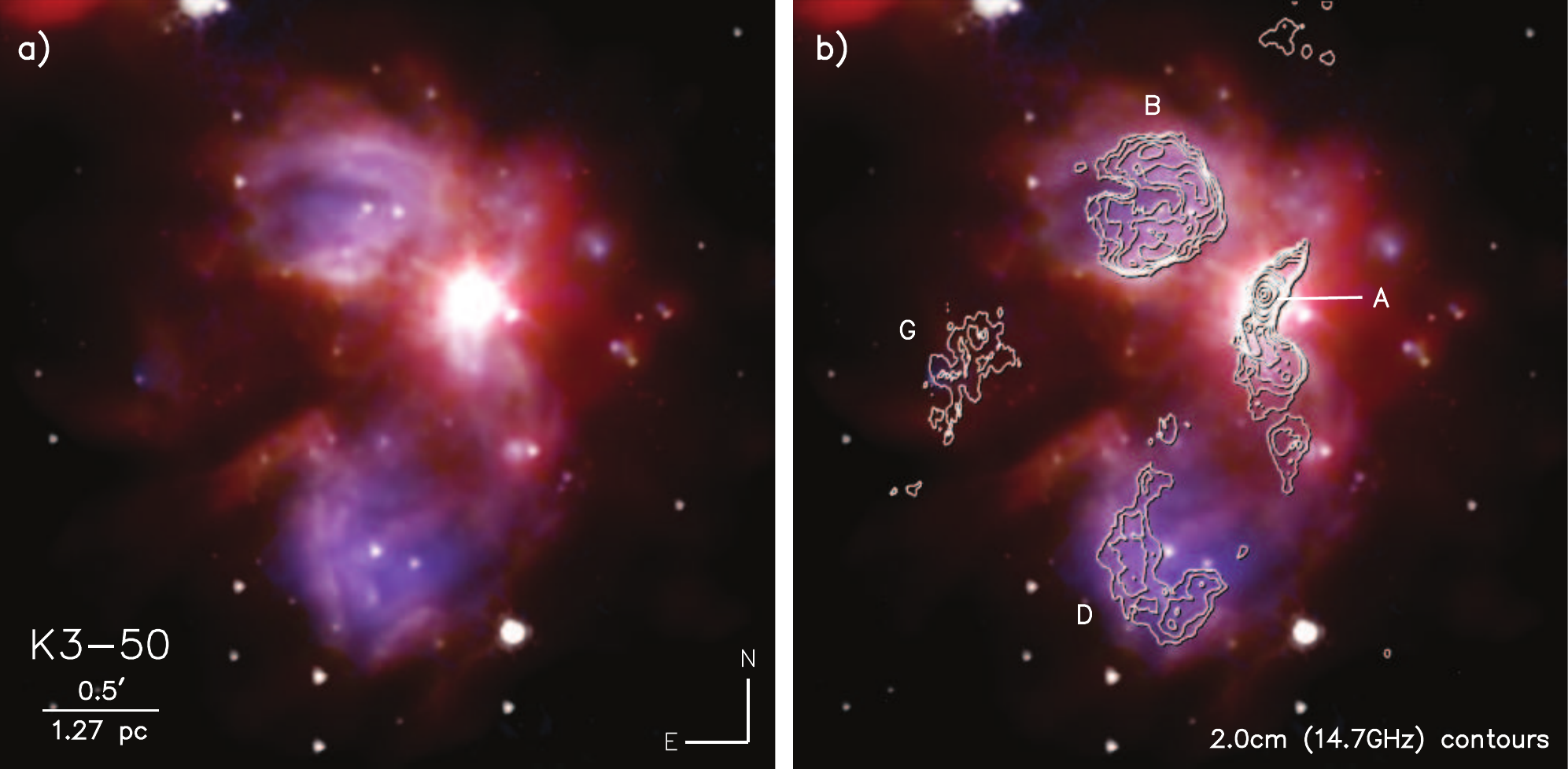}
\caption{Images of K3-50. a) A 3-color image of a $\sim3\farcm0\times 3\farcm$0 field centered on K3-50 where blue is the SOFIA-FORCAST 20\,$\mu$m image, green is the SOFIA-FORCAST 37\,$\mu$m image, and red is the Herschel-PACS 70\,$\mu$m image. Overlaid in white is the Spitzer-IRAC 3.6\,$\mu$m image, which traces the revealed stars within K3-50, field stars, and hot dust. b) Same image overlaid with the 2\,cm (14.7\,GHz) radio continuum contours of \citet{1994ApJ...428..670D} with the radio sub-regions labeled.\label{fig:fig3}}
\end{figure*}

\begin{figure*}[tb!]
\epsscale{1.15}
\plotone{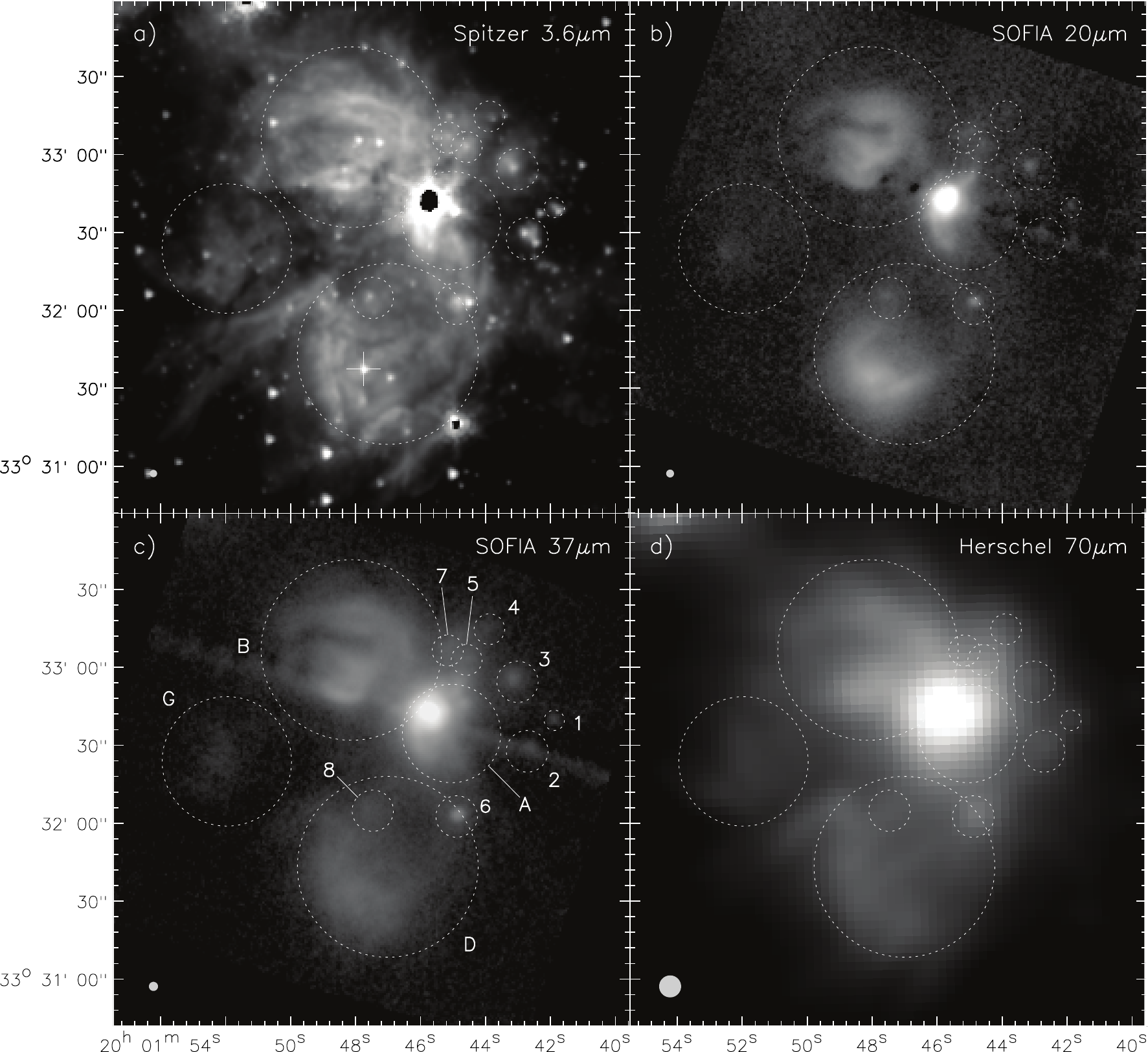}
\caption{Images of K3-50 at a) Spitzer-IRAC 3.6\,$\mu$m, where saturated areas in the center of some stars are black and the location of the central star of K3-50D is marked with a cross, b) SOFIA-FORCAST 20\,$\mu$m, c) SOFIA-FORCAST 37\,$\mu$m, and d) Herschel-PACS 70\,$\mu$m. The line of emission at a p.a. of $\sim$60$\arcdeg$ going through source A and across the entire array in panels b and c is an array artifact and not real dust emission associated with K3-50. The infrared sources are numbered and the sizes and locations of the circles around them corresponds to the coordinates and apertures used for photometry at 37\,$\mu$m (Table \ref{tb:K350_sources}). The gray dot in the lower left corner of each panel shows the spatial resolution at the given wavelength.\label{fig:fig4}}
\end{figure*}

\begin{figure*}[tb!]
\epsscale{1.15}
\plotone{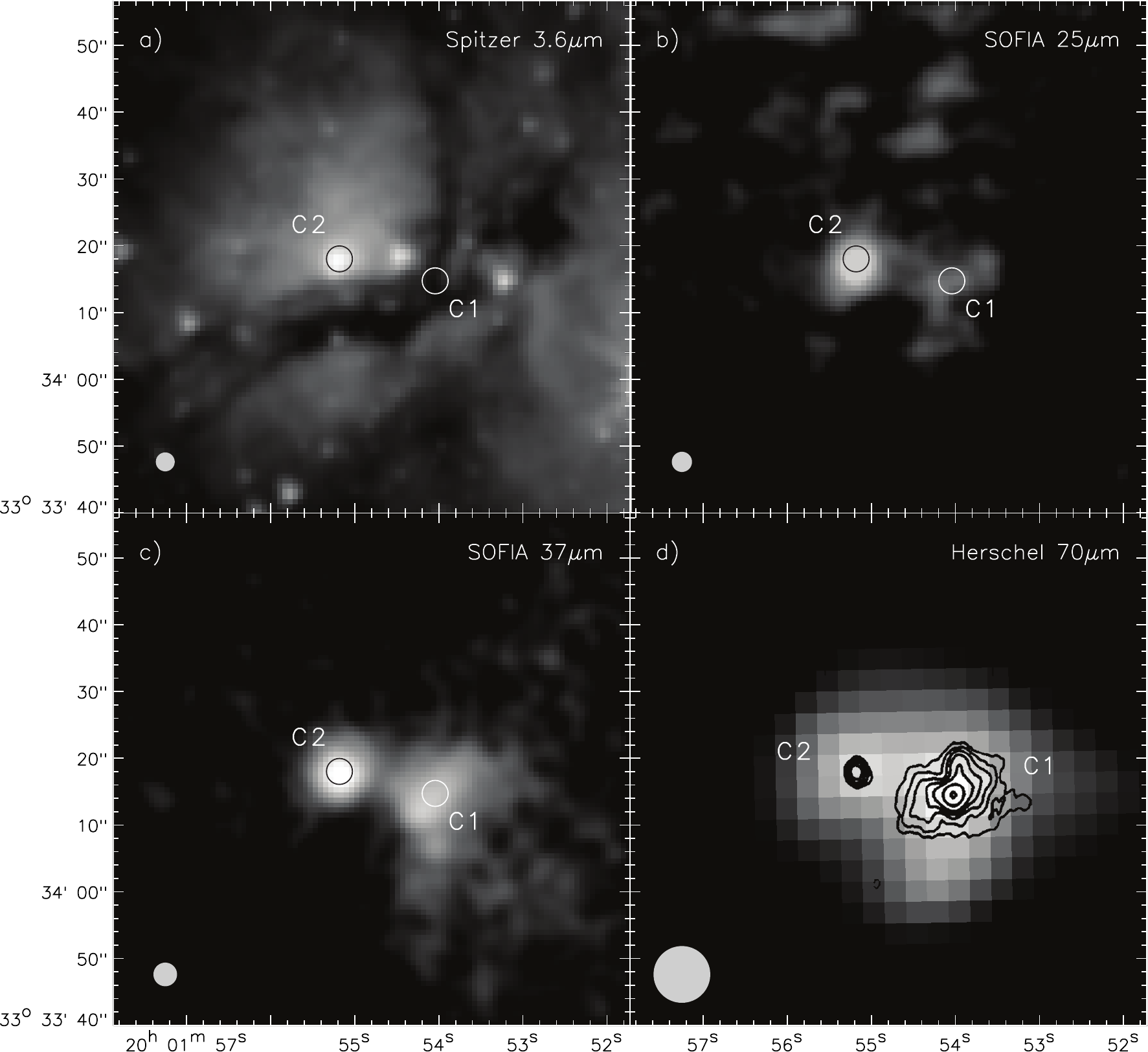}
\caption{Images of region K3-50C at a) Spitzer-IRAC 3.6\,$\mu$m, b) smoothed SOFIA-FORCAST 25\,$\mu$m, c) smoothed SOFIA-FORCAST 37\,$\mu$m, and d) Herschel 70\,$\mu$m. The SOFIA data are smoothed with a 2-pixel Gaussian because of the lower S/N due to short total exposure times. In panel d, shown in black are the  2\,cm radio continuum contours of \citet{1994ApJ...428..670D} overlaid. In panels a-c, the circles show the peak cm radio continuum location of the sources C1 and C2. The resolution of each image is given by the gray dot in the lower left corner of each panel.\label{fig:fig5}}
\end{figure*}

We present images from our SOFIA data of the regions including K3-50 A, B, and D in Figures \ref{fig:fig3} and \ref{fig:fig4}. As discussed in Section~\ref{sec:obs}, we have additional data on region C, but the data were taken with a much shorter exposure time in this region (40\,s on-source) and the shorter wavelength SOFIA data were taken with a 25\,$\mu$m (not 20\,$\mu$m) filter. Therefore, the data for the field containing source C were not mosaicked together with the other K3-50 data. We present the SOFIA images for region C separately in Figure \ref{fig:fig5}. Our SOFIA data do not cover the weaker sources E or F. Since the individual regions of K3-50 are often studied separately (unlike the sources within DR7, which typically are studied together), we will discuss prior observations and our SOFIA results for each region separately below.

\textbf{K3-50 A --} Source A has garnered most of the attention in studies of this region. It is an extremely bright compact \ion{H}{2} region (5.8\,Jy at 2\,cm with peak EM=5.8$\times$10$^8$\,pc\,cm$^{-6}$; \citealt{1994ApJ...428..670D}), and its nature is already well-characterized. It was first shown by \citet{1994ApJ...428..670D} to have an ionized bipolar outflow situated approximately north-south ($p.a.=-25\arcdeg$ east of north). \citet{1997ApJ...477..738H} find the outflow is blue-shifted to the south and therefore has much less extinction than the northern red-shifted outflow lobe. Additionally, it is surrounded by a large molecular toroid (8$\arcsec$ diameter, or 0.3\,pc) that is situated approximately perpendicular to the outflow (disk plane $p.a.\sim$ 55$\arcdeg$), as seen in HCO$^+$ ($J=1-0$) map by \citet{1997ApJ...477..738H}. Rather than being excited by a single massive star, at the center of K3-50 A there is apparently a small cluster of massive and intermediate mass stars, with at least 8 point-sources seen here in the central 3$\arcsec\times$3$\arcsec$ area in the near and mid-infrared \citep{1996ApJ...460..744H, 2003ApJ...584..368O, 2004ApJS..155..123A, 2004A&A...417..981H}. \citet{2003ApJ...584..368O} claim from their mid-infrared observations that at least two, but maybe three, of these sources are massive enough to be ionizing sources powering the C\ion{H}{2} region.

Given the spatial resolution of our SOFIA data, we see only the larger-scale features of this source, namely that it has a very bright core at both 20 and 37\,$\mu$m, and faint extended emission can be seen associated with the outflow both to the north and south (Figure \ref{fig:fig4}b and \ref{fig:fig4}c). The outflow is more prominent at 20\,$\mu$m than 37\,$\mu$m, due to a decrease in flux in the outflow and an increase in the brightness of the core at 37\,$\mu$m. The southern lobe of the outflow is brighter at both wavelengths, as would be expected if it is the blue-shifted lobe. Such mid-infrared signatures of outflow are similar to what has been seen in previous studies of MYSOs
\citep[e.g.,][]{2005ApJ...628L.151D,2006ApJ...642L..57D,2017ApJ...843...33D}.  The southern lobe is also more prominent in the Spitzer-IRAC images as well, though the northern lobe is not seen in the 4.5 or 3.6\,$\mu$m images (Figure \ref{fig:fig4}a).
 
\begin{figure*}[tb!]
\epsscale{1.15}
\plotone{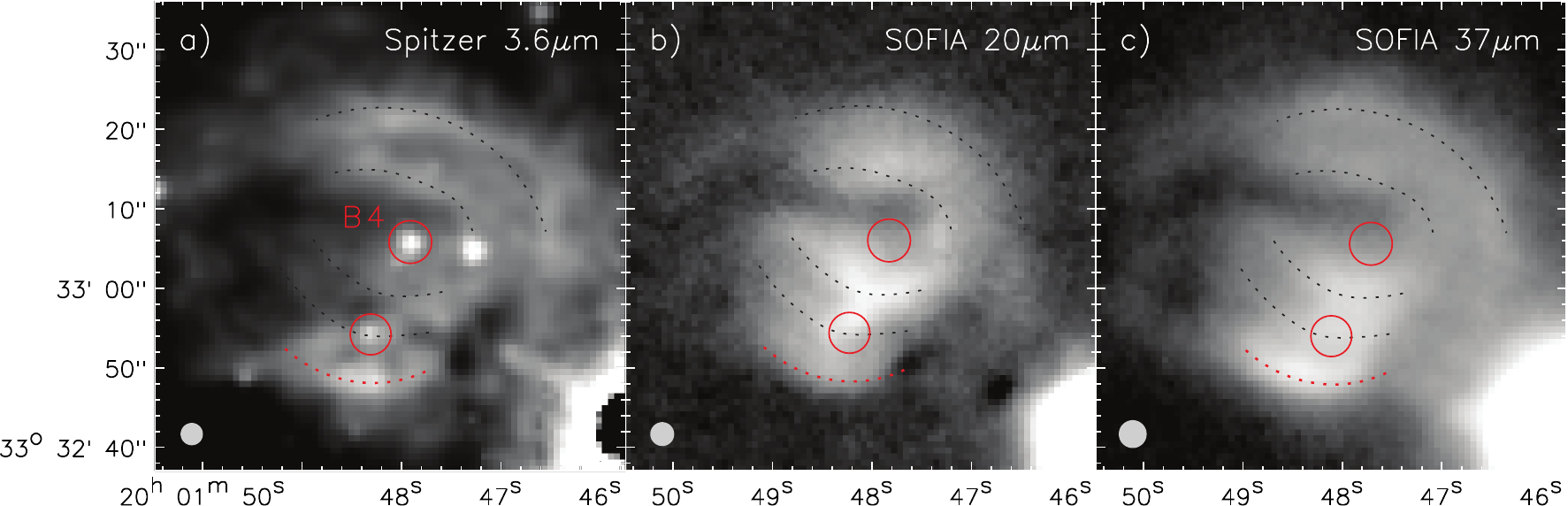}
\caption{Images of region K3-50B at a) Spitzer-IRAC 3.6\,$\mu$m, b) SOFIA-FORCAST 20\,$\mu$m, and c) SOFIA-FORCAST 37\,$\mu$m. The red circles show the locations of the stellar source B4, as well as the stellar source likely responsible for the southern most arc of emission. There is a  double arc structure concentric upon the northern star, and they are outlined by the dashed black lines. A separate arc structure concentric with the southern star is outlined in red. The saturated source in the bottom right is K3-50A. The resolution of each image is given by the gray dot in the lower left corner of each panel. \label{fig:fig6}}
\end{figure*}

\textbf{K3-50 B --} \citet{2010ApJ...714.1015S} suggest this source has a blister-type (a.k.a champagne flow) \ion{H}{2} region morphology based upon $\sim$5$\arcsec$ resolution radio maps at 610\,MHz (49\,cm), with a sharper fall-off in emission to the west than east.  However, at around 1$\arcsec$ spatial resolution and at 2\,cm wavelengths \citep[e.g.,][]{1994ApJ...428..670D}, this sharper western fall-off is not really apparent, but instead it appears as an internally flocculent structure whose overall morphology resembles a backwards ``c''. \citet{1994ApJ...428..670D} described it as having an incomplete shell morphology, with a gap towards the east. This gap is not only seen in the radio (Figure \ref{fig:fig3}b), but is present in the infrared from 3\,$\mu$m all the way out to 20 and 37\,$\mu$m (as seen in the SOFIA images), and can even be seen in the Herschel 70\,$\mu$m data (Figure \ref{fig:fig4}), indicating a relative lack of gas or dust in the eastern part of the source (i.e., rather than being due to high levels of extinction). Located nearly in the center of the B region is a near-infrared-bright star (Figure~\ref{fig:fig6}a) which is an O5 ZAMS star named B4 by \citet{2010ApJ...714.1015S}, that they claim may be solely responsible for clearing the inner hole and heating and ionizing the entire partial gas and dust shell (Figure \ref{fig:fig6}a).

As discussed by \citet{1996ApJ...460..744H}, source B appears to have a double shell structure at 3.29\,$\mu$m, and conjecture that is may be due to episodic mass loss associated with star formation. Comparable double shell morphologies have been seen in our larger G\ion{H}{2} region study, specifically around sources G49.4-0.3a and  G49.4-0.3c in W51A (\citetalias{2019ApJ...873...51L}). However, the W51A sources are similar to what is seen in our SOFIA data of K3-50 B (Figure~\ref{fig:fig6}), with an important distinction. At 37\,$\mu$m, there are two arcs of emission 10$\arcsec$ and 14$\arcsec$ away from the location of the star B4, both to the north of the star and to the south (Figure~\ref{fig:fig6}c). At 20 \,$\mu$m the northern two arcs blend into a single broad one, but the southern pair can be seen as separate structures (Figure~\ref{fig:fig6}b). However, further to the south lies a third partial shell arc that is not replicated to the north. It is apparent in the Spitzer IRAC data as well as the SOFIA data. Located about 5$\arcsec$ north of the apex of this arc lies another NIR stellar source most readily seen in the Spitzer 3\,$\mu$m data (Figure~\ref{fig:fig6}a). Given its location within the arc and that the arc is extended to the east the same amount as to the west of this NIR star, it is likely that this star is responsible for heating this third, southern-most dust arc. Interestingly, this stellar source we believe is heating this southern-most arc is located within the larger second arc around B4. This could just be a coincidence of projection, however it could also be that the swept-up material in the southern part of the secondary ring collapsed to form this star. 

Too big to be considered a compact \ion{H}{2} region \citep[$\sim$0.1 $<d<$ 0.5\,pc;][]{1967ApJ...150..807M}, K3-50 B would instead be considered to be an expanding \ion{H}{2} region with its measured diameter of $\sim$1.2\,pc. The B region is brighter and more compact at 20\,$\mu$m than 37\,$\mu$m, suggesting that all of the dust here is internally heated (by the central O star, B4, presumably) and the markedly ``blue'' interior in the 3-color image (Figure \ref{fig:fig3}) is due to the inside being hotter (and thus brighter at 20\,$\mu$m) than the outside of the region. 

\textbf{K3-50 C --} \citet{1975MNRAS.170..139H} resolved K3-50 C into two sources (C1 and C2) at 6\,cm separated by $\sim$15$\arcsec$ with C2 appearing more compact (with only a slight north-south elongation), and C1 having a peak with emission extending out to 5$\arcsec$ from the core in the NW and SE directions. \citet{1994ApJ...428..670D} describe C1 as having more of a ``core-halo’’ C\ion{H}{2} morphology at 2\,cm, while \citet{1996ApJ...460..744H} discuss that the extension is due to an outflow of ionized gas. In the near-infrared, the C region appears as an extended emission area with $r\sim20\arcsec$ and with the radio C1 and C2 peaks near its center (Figure\,\ref{fig:fig5}a). Clearly visible in the Spitzer-IRAC images (as well as the H and K images of \citealt{1996ApJ...460..744H}), is a dark dust lane running more or less east-west across the extended emission of the C region, centered on the radio continuum peak of C1. \citet{1988A&A...207..132R} postulate that this could be an edge-on molecular disk around C1. The radio peak of C2 lies about 8$\arcsec$ north of the mid-plane of the dark lane. The presence of this dust lane is likely contributing to the rather large estimates of visual extinction towards C1 and C2 ($\ge$190 mag and 32 mag, respectively) derived by \citet{1988A&A...207..132R}. Additionally, the $^{12}$CO ($J=1-0$) maps of \citet{1980ApJ...236..465I} at $\sim$2$\arcmin$ resolution show that the peak concentration of molecular gas in the K3-50 region is located at region C, which again explains the high observed extinction there.

\citet{1977MNRAS.179..255W} claim that C2 is more prominent between 2 and 20\,$\mu$m (however, C1 is stronger than C2 at 8.4\,GHz; \citealt{1994ApJS...91..659K}). C1 was only detected with certainty (but barely) at 25\,$\mu$m by \citet{1977MNRAS.179..255W}, but not at shorter wavelengths. In our SOFIA data we can confirm that the emission from C2 is unresolved and is indeed much more prominent than C1 at 25\,$\mu$m (see Figure~\ref{fig:fig5}). At all infrared wavelengths from 3-70\,$\mu$m (i.e., Spitzer, SOFIA, and Herschel) the peak of C2 in the infrared appears compact and is coincident with the radio peak of C2, and thus this source might house a single MYSO. 

Like \citet{1977MNRAS.179..255W}, at 25\,$\mu$m we detect diffuse and extended emission at the C1 location, but with no real defined peak (Figure\,\ref{fig:fig5}). However, at 37\,$\mu$m, C1 is much brighter than at 20\,$\mu$m and more extended as well. The overall extension is northwest to southeast, similar to what is seen in the radio continuum images  \cite[e.g.,][]{1994ApJ...428..670D}. We still do not detect a clearly defined peak in the infrared emission at 37\,$\mu$m for C1, but at this wavelength the integrated flux of C1 is now greater than that of the unresolved C2 point source. At 70\,$\mu$m, C1 appears in the Herschel data to have a peak coincident with the cm radio continuum peak location, and this peak is much brighter than the peak of C2. C1 also can be seen in the Herschel 70\,$\mu$m data to have a clear elongation from the northwest to southeast. Since \citet{1996ApJ...460..744H} discuss the elongated radio continuum emission of C1 as potentially tracing a partially ionized outflow, the infrared extension of emission we are seeing in the SOFIA and Hershel data may be the dust cavity walls or the dust in the outflow, as has been seen in many MYSO outflow sources \citep[e.g.,][]{2006ApJ...642L..57D, 2017ApJ...843...33D}. 

\textbf{K3-50 D --} Source D corresponds to the radio continuum emission coming from the optical emission nebula NGC 6857.  The optical emission from this region is brightest around an optical-IR star at $\alpha_{J2000} = 20^h01^m47.7^s$, $\delta_{J2000} = +33\arcdeg31\arcmin37.4\arcsec$ (see cross in Figure\,\ref{fig:fig4}a), which was identified by \citet{2010ApJ...714.1015S} as a ZAMS O5-O6 star, and is believed to be the source ionizing and heating the entire D region. In the optical there is a fan of dim and diffuse emission extending for an additional $\sim$3$\arcmin$ to the southwest. In the NIR ($I$ through $K$ bands), the emission appears to be much more compact, taking on the shape of a partial shell ($r\sim17\arcsec$) around the central O star, with an opening to the northwest in the direction of K3-50 A. At Spitzer-IRAC wavelengths, rather than a simple shell structure, the region is more extended and filled with flocculent sub-structure and ridges creating an overall V-shape appearance pointing to the south-southeast. 

The SOFIA 20 and 37\,$\mu$m images appear much more similar in extent and shape (i.e., arc or partial shell structure) to the $I-K$ band images and appear very similar in morphology to the cm radio continuum emission (see Figure~\ref{fig:fig3}). 

\begin{deluxetable*}{rllccccccl}
\tabletypesize{\scriptsize}
\tablecolumns{10}
\tablewidth{0pt}
\tablecaption{SOFIA-FORCAST Observational Parameters of Infrared Sources in K3-50}
\tablehead{\colhead{ }&
           \colhead{  } &
           \colhead{  } &
           \multicolumn{3}{c}{${\rm 20\mu{m}}$}&
           \multicolumn{3}{c}{${\rm 37\mu{m}}$} &
           \colhead{}\\
           \cmidrule(lr){4-6} \cmidrule(lr){7-9}\\
           \colhead{ Source }&
           \colhead{ R.A.(J2000) } &
           \colhead{ Dec.(J2000) } &
           \colhead{ $R_{\rm int}$ } &
           \colhead{ $F_{\rm int}$ } &
           \colhead{ $F_{\rm int-bg}$ } &
           \colhead{ $R_{\rm int}$ } &
           \colhead{ $F_{\rm int}$ } &
           \colhead{ $F_{\rm int-bg}$ } &
           \colhead{Aliases}\\
	   \colhead{  } &
	   \colhead{  } &
	   \colhead{  } &
	   \colhead{ ($\arcsec$) } &
	   \colhead{ (Jy) } &
	  \colhead{ (Jy) } &
	   \colhead{ ($\arcsec$) } &
	   \colhead{ (Jy) } &
	   \colhead{ (Jy) } &
	   \colhead{}  
}
\startdata
 \multicolumn{10}{c}{Compact Sources} \\
 \hline
 K3-50 1 & 20:01:41.8     & +33:32:39.7     &5	   		&0.63       &0.60		&5		&2.04	&1.67  &   \\
 K3-50 2 & 20:01:42.7     & +33:32:27.8     &6         &$<$1.36$^d$       &\nodata 	&8    	&10.4	&5.11  &   \\
 K3-50 3 & 20:01:43.0     & +33:32:54.5     &7	   		&4.41       &2.66		&8		&21.9	&12.8  &   \\ 
 K3-50 4 & 20:01:43.8     & +33:33:14.8     &5    		&1.67       &0.33 		&6    	&10.3	&4.74  &   \\
 K3-50 5 & 20:01:44.5     & +33:33:03.1     &5	   		&3.45       &1.58		&6		&$<$37.8$^e$	&\nodata  &   \\
 K3-50 6 & 20:01:44.8     & +33:32:02.6     &6	   		&6.29       &3.38 		&8		&40.0	&22.1  &   \\
 K3-50 7 & 20:01:45.1     & +33:33:06.6     &6    		&5.70       &2.33 		&6  	&33.8	&10.3 &   \\
 K3-50 8 & 20:01:47.5     & +33:32:04.6     &8    		&9.40       &3.55 		&8  	&29.4	&13.3 &   \\
 K3-50 C1 & 20:01:54.1 	   & +33:34:14.7 	 &13	   	&{\it 76.5}$^c$       &{\it 46.0}$^c$		&13		&406	&362  &   \\	   
 K3-50 C2 & 20:01:55.2     & +33:34:18.0 	 &8    	&{\it 51.8}$^c$       &{\it 51.4}$^c$		&8    	&223	&221  &   \\
 \hline
 \multicolumn{10}{c}{Extended Sources} \\
 \hline
 K3-50 A & 20:01:45.0$^a$ & +33:32:34.6$^a$ &18   		&985        &956 		&19   	&4280	&4050  &   \\
 K3-50 B & 20:01:48.1$^b$ & +33:33:06.8$^b$ &23   		&198        &165 		&35		&1140	&941  &  \\
 K3-50 C & 20:01:54.6$^b$ & +33:34:16.0$^b$ &23   		&{\it179}$^c$        &{\it 117}$^c$ 		&31		&926	&913  &  \\
 K3-50 D & 20:01:47.0$^b$ & +33:31:43.3$^b$ &31   		&282        &253 		&35    	&577	&471  &  NGC\,6857 \\	   
 K3-50 G & 20:01:52.0$^b$ & +33:32:23.8$^b$ &25   		&12.0       &4.52 		&25		&22.8	&21.0  &   		   	   
\enddata
\tablecomments{\footnotesize R.A. and Dec. are for the center of the photometric apertures used. $R_{\rm int}$ indicates radius of the aperture. $F_{\rm int}$ is the integrated flux in the aperture, while $F_{\rm int-bg}$ is that same flux with an estimate of the surrounding background subtracted. See Section \ref{sec:data} for more information.}
\tablenotetext{a}{The aperture had to be offset to encompass all of the flux from source A without contamination for source B; actual mid-infrared peak location is at $\alpha_{J2000}$=20:01:45.7, $\delta_{J2000}$=+33:32:42.1.}
\tablenotetext{b}{Sources B, C1, D, and G are essentially peak-less, and so the coordinates in the table are for the center of the aperture used which best encompasses all of the flux from each source, and does not represent to center or peak of their flux distributions. Similarly the coordinates for C are the center of the 37\,$\mu$m emission of the region.}
\tablenotetext{c}{The italicized text means that sources C, C1, and C2 were taken with the 25\,$\mu$m filter instead of the 20\,$\mu$m filter. The photometric uncertainty is higher in this filter (25\%).}
\tablenotetext{d}{There are instrument array artifacts that go through this source add an unknown quantity of excess emission and therefore this value is considered an upper limit.}
\tablenotetext{e}{This source is not fully resolved from the much brighter source A at 37\,$\mu$m and therefore this value is considered an upper limit.}
\label{tb:K350_sources}
\end{deluxetable*}

The 23\,cm radio maps of \citet{2010ApJ...714.1015S} show best that the arc of the D region extends on the western side all the way to source A. The overall impression one gets is a hook or j-shaped morphology. While the emission is brightest in the southern-most arc of the source, we do see faint infrared emission in the SOFIA 20 and 37\,$\mu$m images tracing this ``bridge'' seen in cm radio continuum emission towards source A. Embedded in this bridge of emission about halfway between A and D lies a very bright infrared source in our SOFIA images (Figure~\ref{fig:fig4}), which we name source 6, that can also be easily seen in the Spitzer-IRAC images. There is also a knot of cm radio continuum emission nearby, but not quite coincident, with the source in the 2\,cm images of \citet{1994ApJ...428..670D}. Our SED model fitting (discussed in Section~\ref{sec:cps}) show that this source is most likely a very massive YSO (best model fit of $M_{star}=48M_{\sun}$). We see a second source at the tip of the hook or ``j'' (Figure~\ref{fig:fig4}). This source, which we call source 8, is more apparent in the SOFIA 20\,$\mu$m image than the 37\,$\mu$m image, but there is a peak at this location in all Spitzer-IRAC bands (but it does not have an optical component). Again there is also a knot of 2\,cm radio continuum emission nearby, but not quite coincident with this location in the images of \citet{1994ApJ...428..670D}. Our SED model fitting to this source identify it as an intermediate to high-mass YSO ($4 <M_{star}<16M_{\sun}$; see Section~\ref{sec:cps}).  

The fact that the gap in the shell of infrared and radio emission lies on the side facing source A, and that the outflow from source A is towards source D, this might indicate that the outflow from A may be influencing the morphology of source D. We caution that this is based upon morphologies of 2-dimensional projected images, and kinematics of the gas would be needed to know if this was the case with any certainty. 

Because this is the only extended emission region in K3-50 that is optically visible (with measured A$_V\sim$ 2\,mags; \citealt{1988A&A...207..132R}) this region is likely to be the most evolved sub-region of K3-50. Therefore, like source B, the shell of gas and dust we see here in the radio and infrared is the remains of a cavity carved by a massive O star. Since a compact \ion{H}{2} region is defined to have a maximum size near $d\sim$ 0.5\,pc \citep{1967ApJ...150..807M}, at a size of $d\sim$ 1.4\,pc, K3-50 D would be considered an expanding \ion{H}{2} region. Given the changes in morphology and appearance as a function of wavelength, the \ion{H}{2} region may have been expanding through a rather clumpy environment. Like the B region, the D region is brighter and more compact at 20\,$\mu$m than 37\,$\mu$m, again pointing to the dust here being dominantly heated by the central O star. Also like region B, interior of the D region is very ``blue'' in the 3-color image (Figure~\ref{fig:fig3}), showing that the 20\,$\mu$m emission is tracing hotter dust closer to the star and the 37 and 70\,$\mu$m emission is tracing the cooler dust in the outer parts of the \ion{H}{2} region shell.

\textbf{New region: K3-50G --}  About 80$\arcsec$ east of source A, there lies a diffuse patch of 2\,cm radio continuum emission (Figure\,\ref{fig:fig3}b), as seen in the images of \citet{1994ApJ...428..670D}. The extent of the radio continuum emission is about $\sim$20$\arcsec$ across, and we see a similarly-sized patch of diffuse emission in our 20 and 37\,$\mu$m images (Figure~\ref{fig:fig4}). The region is also visible as an area of emission separated from the extended emission of K3-50 A, B, and D by a dark arc in the Spitzer-IRAC images, and is also discernible as a fainter source in the Herschel 70\,$\mu$m data. The Spitzer images are sensitive enough to reveal the partial shell morphology of the source ($r\sim5\arcsec$), with an opening to the northeast (Figure~\ref{fig:fig4}a). The Spitzer images also reveal the presence of a stellar source, located near the center of this shell. This star is also seen in the Digitized Sky Survey $I$-band images. This region is also identified as a source in the MSX Point Source Catalog Version 2.3 \citep{2003yCat.5114....0E} and is named MSX6C:G070.2990+01.5762. Coincident with the stellar location there is a slight peak in the SOFIA 20\,$\mu$m image, but the 37\,$\mu$m image shows only a broad diffuse structure. The emitting region at 37\,$\mu$m is displaced slightly to the west of the 20\,$\mu$m emitting region and the Herschel 70\,$\mu$m emission appears as an arc slightly further to the west of the 37\,$\mu$m emitting region. Since the shorter wavelength infrared emission is closer to the location of the stellar source MSX6C:G070.2990+01.5762, this star may be responsible for heating this region. Interestingly, the radio continuum is only associated with the westernmost part of the partial shell and some radio emission extends further to the northwest than the dust emission. Given that the region is pervaded with cm radio continuum emission, the stellar source must be massive enough to ionize the region as well. This means that this new source, which we dub K3-50 G, may be a fainter analog to the more evolved, partial shell sources K3-50 B and K3-50 D with revealed near-infrared central ionizing and heating sources.

\section{Data Analysis and Results}\label{sec:data}

All infrared sources identified in Section \ref{sec:results1} are tabulated in Table \ref{tb:DR7_sources}, in the case of DR7, and Table \ref{tb:K350_sources}, in the case of K3-50. In these tables we specify the right ascension and declination of the aperture centers (which are sometimes, but not always, the source peaks or centers) used for the photometry of each source as well as the aperture radii used at each wavelength ($R$\textsubscript{int}). We give the integrated flux densities at both wavelengths within those apertures ($F$\textsubscript{int}), as well as background subtracted estimates of the flux densities of all sources. We apply the same aperture photometry practices as we did in our previous studies to ascertain the aperture sizes to use for flux extraction. To quickly summarize, we choose an aperture radius where the flux from the azimuthally-averaged radial profile of a source just begins to level out. If the source is surrounded by extended emission, this background is only a local minimum. The background flux estimate is taken from the statistics of the data within an annulus just outside that aperture, the thickness of which is determined by the range of radii where the background remains at a constant level. These background subtracted flux estimates are given in the tables in the columns labeled $F$\textsubscript{int-bg}. 

Of the ten sources identified in the infrared for DR7, five are newly identified for the first time here. On the other hand, two radio-defined sources in DR7, D and E, do not seem to have prominent infrared counterparts (though extended infrared emission lies throughout the areas of both radio sources). For K3-50 we identified 14 sources (when including C/C1/C2), of which eight are newly identified sources. We detect all previously identified radio continuum sources in K3-50 that were contained in the region covered by the SOFIA fields.

\subsection{Physical Properties of Compact Sources: SED Model Fitting and Determining MYSO Candidates}\label{sec:cps}

\begin{figure}
\center
\begin{tabular}[b]{c@{\hspace{-0.1in}}c}
\includegraphics[width=3.2in]{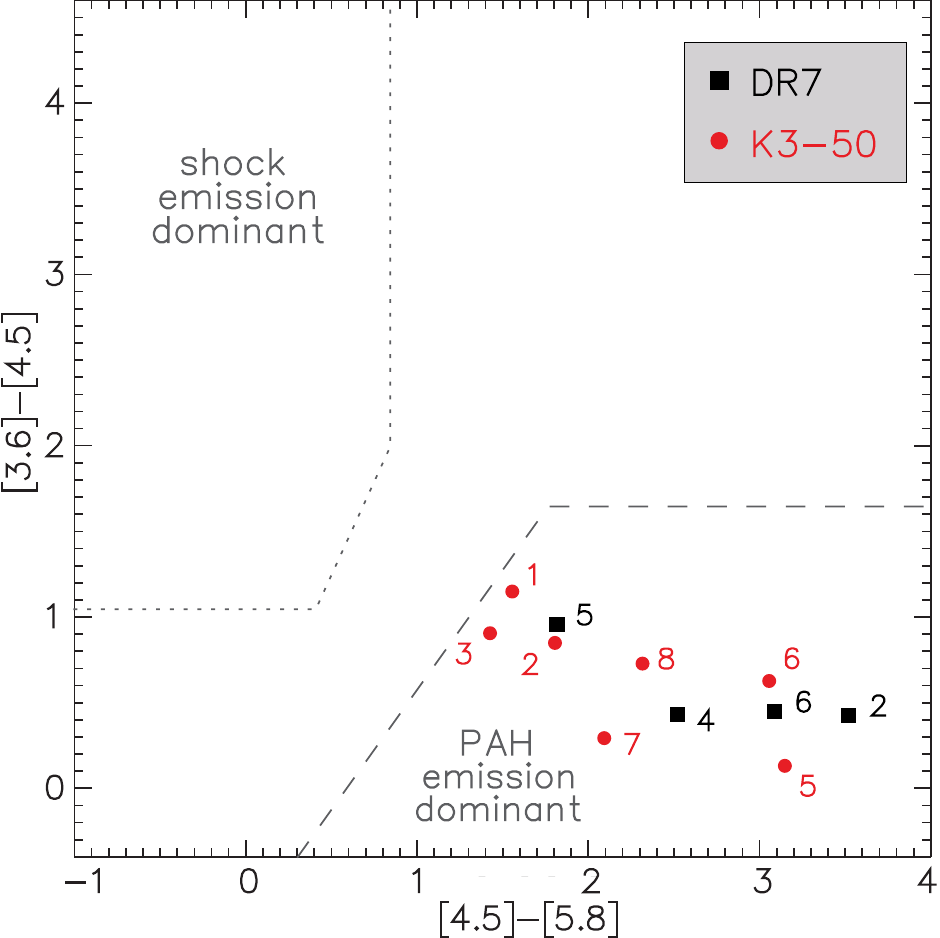}\\
\end{tabular}
\caption{\footnotesize A color-color diagram utilizing our background-subtracted Spitzer-IRAC 3.6, 4.5, and 5.8\,$\mu$m  photometry to distinguish which of the identified compact sources are ``shocked emission dominant'' and ``PAH emission dominant''. Above (up-left) of the dotted line indicates shock emission dominant regime. Below (bottom-right) the dashed line indicates PAH dominant regime. We adopt this metric from \citet{2009ApJS..184...18G}. The black squares are the sources within DR7 and source names are labeled. The red dots are the sources within K3-50. Sources K3-50 4, C1, and C2 are not included in this diagram due to non-detection or saturation in the Spitzer-IRAC bands.}\label{fig:ccd}
\end{figure}

We further subdivide the sources for each region into whether they are compact or extended sources. The two categories denote which objects we believe are star-forming cores (compact) vs. the larger star-forming molecular clumps (extended). We make this distinction so that we can isolate the star-forming cores that we wish to apply MYSO SED models to, since such models are not meant for fitting molecular clumps or clouds. We further define a compact source as one that has a definitive peak that does not change location significantly with wavelength. To count as a source, it must also be detected at more than one wavelength. 

Given the comparable distances to both DR7 and K3-50, we will use the same angular size criteria to select the sub-sample of compact sources. We will consider any source where we employed a photometric aperture radius of $\le$8$\arcsec$ ($\lesssim0.3$\,pc, which is similar in physical size to the compact sources in our previous studies) in Tables \ref{tb:DR7_sources} and \ref{tb:K350_sources} as a compact source. This means the compact sources in DR7 are sources 2, 4, 5 and 6, and in K3-50 are sources 1 through 8 and source C2. Though the aperture we use for the photometry of C1 is larger than 8$\arcsec$, we make an exception and include this source in compact category as well, since we will treat the entirety of the C region in our study on the extended regions within K3-50.

\begin{figure*}[tb!]
\epsscale{1.10}
\plotone{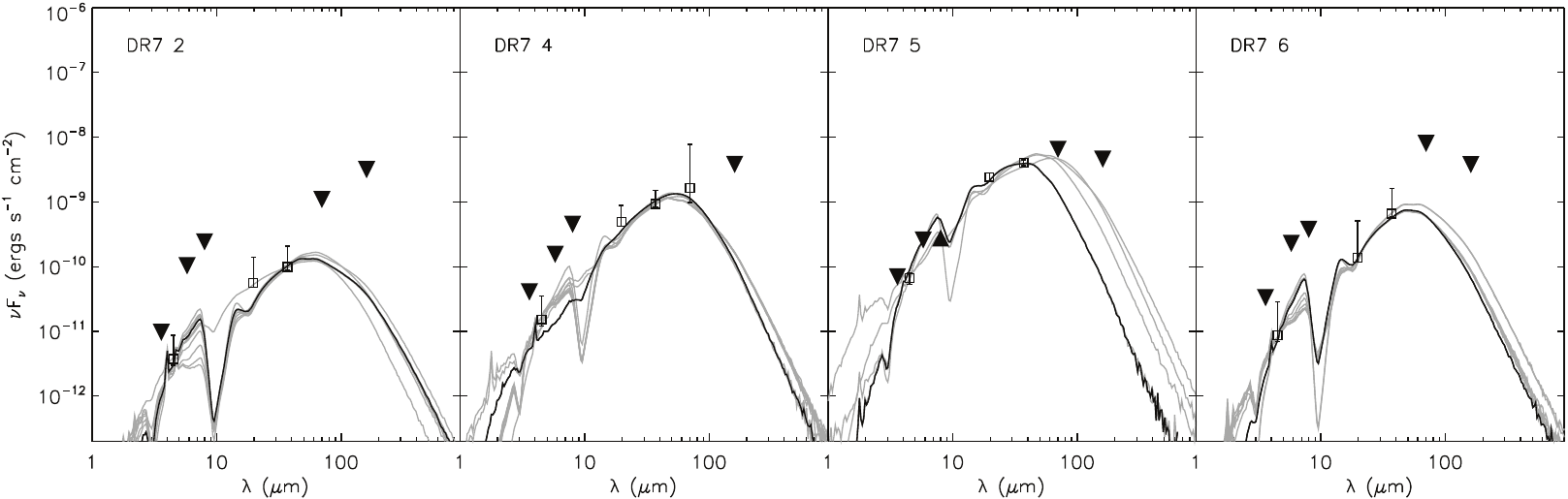}
\caption{SED fitting with ZT model for compact sources in DR7. Black lines are the best fit model to the SEDs, and the system of gray lines are the remaining fits in the group of best fits (from Table \ref{tb:sed_DR7}).  Upside-down triangles are data that are used as upper limits in the SED fits, right-side up triangles are lower limits.\label{fig:DR7SED}}
\end{figure*}

\begin{figure*}[tb!]
\epsscale{1.10}
\plotone{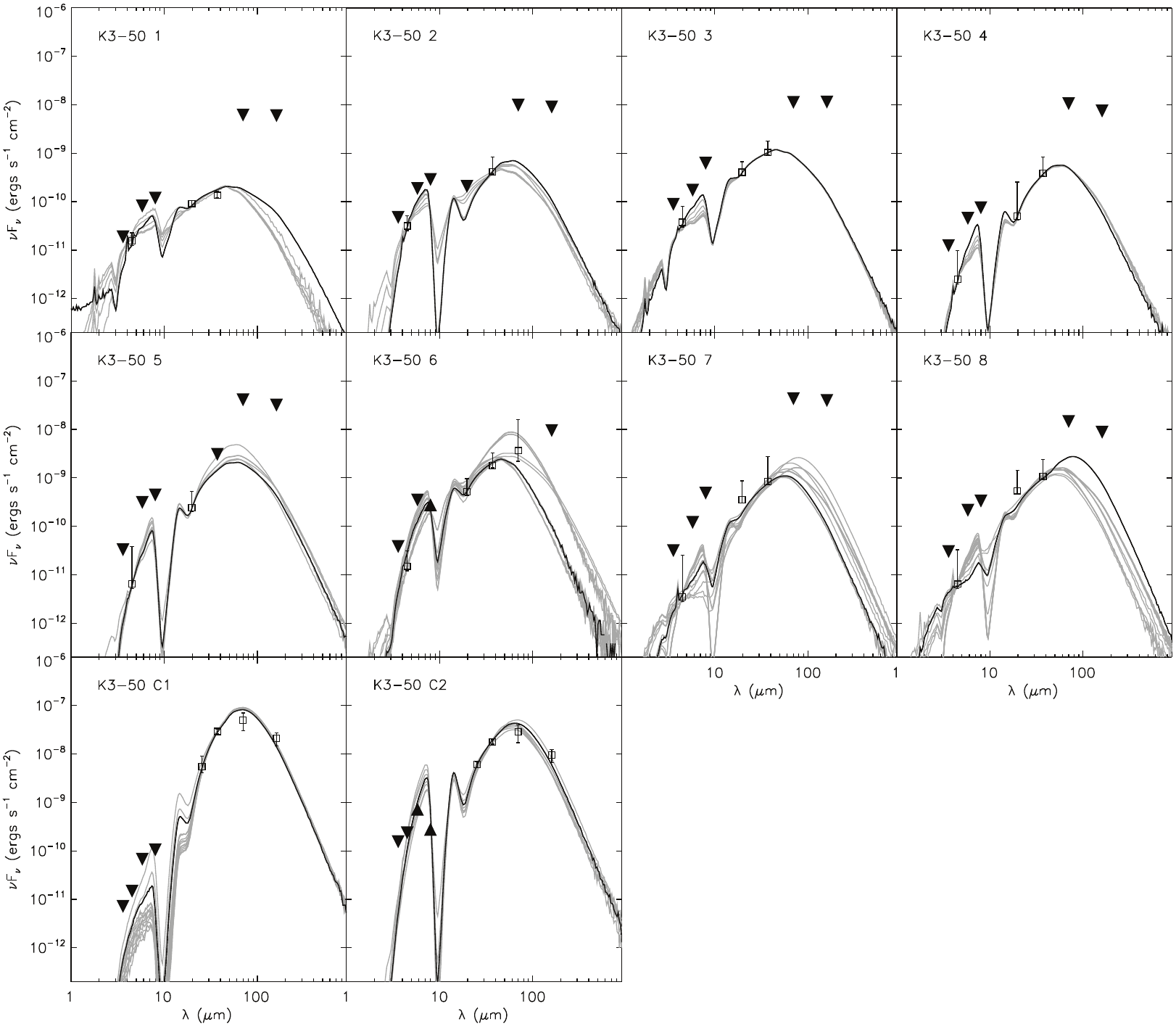}
\caption{SED fitting with ZT model for compact sources in K3-50. Black lines are the best fit model to the SEDs, and the system of gray lines are the remaining fits in the group of best fits (from Table \ref{tb:sed_K350}).  Upside-down triangles are data that are used as upper limits in the SED fits, right-side up triangles are lower limits.\label{fig:K350SED}}
\end{figure*}

In order to create spectral energy distributions (SEDs) for these compact sources, we utilize the SOFIA photometry at 20 and 37\,$\mu$m combined with photometry from the Spitzer and Herschel missions. We performed multi-band aperture photometry on the Spitzer-IRAC 3.6, 4.5, 5.8, 8.0\,$\mu$m data (see Table~\ref{tb:IRAC_DR7} for DR7 and Table~\ref{tb:IRAC_K350} for K3-50) and Herschel-PACS 70 and 160\,$\mu$m data (see Table~\ref{tb:PACS_DR7} for DR7 and Table~\ref{tb:PACS_K350} for K3-50) on all sources. We employed the same optimal extraction technique as in \citetalias{2019ApJ...873...51L} to find the optimal aperture to use for photometry. Background subtraction was also performed in the same way as \citetalias{2019ApJ...873...51L} (i.e. using background statistics from an annulus outside the optimal extraction radius which had the least environmental contamination).

As we did in \citetalias{2019ApJ...873...51L}, we used a color-color ($3.6-4.5$\,$\mu$m vs. $4.5-5.8$\,$\mu$m) plot analysis developed by \citet{2009ApJS..184...18G} to determine which Spitzer-IRAC photometric data may be contaminated with excess flux at 3.6, 5.8, and 8.0\,$\mu$m from polycyclic aeromatic hydrocarbons (PAHs) or at 4.5\,$\mu$m from shocked-excited H$_2$ emission. We show the color-color plot and the data for all sources in Figure~\ref{fig:ccd}. This plot shows that all sources show contamination from PAH emission, and as a consequence all sources will have their Spitzer-IRAC 3.6, 5.8, and 8.0\,$\mu$m photometric values set as upper limits in the SED diagrams; in other words, only the 4.5\,$\mu$m data will be treated as a valid nominal data point. There are some sources missing from the analysis (K3-50 4, C1, and C2) due to non-detection or saturation in the Spitzer-IRAC bands. As we did in \citetalias{2019ApJ...873...51L}, we will treat these sources as average sources, i.e. assume they are “PAH emission dominant”.

\begin{deluxetable*}{rccccrclrclcl}
\tabletypesize{\small}
\tablecolumns{12}
\tablewidth{0pt}
\tablecaption{SED Fitting Parameters of Compact Infrared Sources in DR7}\label{tb:sed_DR7}
\tablehead{\colhead{   Source   }                                              &
           \colhead{  $L_{\rm obs}$   } &
           \colhead{  $L_{\rm tot}$   } &
           \colhead{ $A_v$ } &
           \colhead{  $M_{\rm star}$  } &
           \multicolumn{3}{c}{$A_v$ Range}&
           \multicolumn{3}{c}{$M_{\rm star}$ Range}&
           \colhead{ Best }&
           \colhead{Notes}\\
	   \colhead{        } &
	   \colhead{ ($\times 10^3 L_{\sun}$) } &
	   \colhead{ ($\times 10^3 L_{\sun}$) } &
	   \colhead{ (mag.) } &
	   \colhead{ ($M_{\sun}$) } &
       \multicolumn{3}{c}{(mag.)}&
       \multicolumn{3}{c}{($M_{\sun}$)}&
       \colhead{  Models   } &
       \colhead{   }
}
\startdata
DR7 2 &      0.35 &      0.77 &     19.3 &      4.0 &   2.5 & - &  54.5 &   4.0 & - &   4.0 & 11 & $\dagger$  \\
DR7 4 &      3.01 &     13.30 &     26.5 &      8.0 &   8.4 & - &  79.5 &   8.0 & - &  16.0 & 11 & MYSO\\
DR7 5 &      9.06 &    101.57 &     26.5 &     16.0 &   1.7 & - &  26.5 &   8.0 & - &  16.0 &  5 & MYSO\\
DR7 6 &      1.61 &     11.20 &     58.7 &      8.0 &  10.6 & - &  78.8 &   8.0 & - &  12.0 &  9 & MYSO
\enddata
\tablecomments{\footnotesize  A ``MYSO'' in the right column denotes a MYSO candidate.}
\tablenotetext{\dagger}{\footnotesize The SED fits do not go through the 20\,$\mu$m data point. Therefore the fits are all under-fitting the data (and therefore underestimating the luminosity)}
\end{deluxetable*}

\begin{deluxetable*}{rccccrclrclcl}
\tabletypesize{\small}
\tablecolumns{12}
\tablewidth{0pt}
\tablecaption{SED Fitting Parameters of Compact Infrared Sources in K3-50}\label{tb:sed_K350}
\tablehead{\colhead{   Source   }                                              &
           \colhead{  $L_{\rm obs}$   } &
           \colhead{  $L_{\rm tot}$   } &
           \colhead{ $A_v$ } &
           \colhead{  $M_{\rm star}$  } &
           \multicolumn{3}{c}{$A_v$ Range}&
           \multicolumn{3}{c}{$M_{\rm star}$ Range}&
           \colhead{ Best }&
           \colhead{Notes}\\
	   \colhead{        } &
	   \colhead{ ($\times 10^3 L_{\sun}$) } &
	   \colhead{ ($\times 10^3 L_{\sun}$) } &
	   \colhead{ (mag.) } &
	   \colhead{ ($M_{\sun}$) } &
       \multicolumn{3}{c}{(mag.)}&
       \multicolumn{3}{c}{($M_{\sun}$)}&
       \colhead{  Models   } &
       \colhead{   }
}
\startdata
K3-50 1 &      0.70 &      0.79 &      0.8 &      4.0 &   0.8 & - &  31.8 &   4.0 & - &  32.0 &  6 &      \\
K3-50 2 &      1.71 &     11.66 &    159.0 &      8.0 &  21.0 & - & 159.3 &   8.0 & - &  16.0 &  7 & MYSO \\
K3-50 3 &      3.23 &      9.48 &     12.6 &      8.0 &   2.5 & - &  12.6 &   8.0 & - &   8.0 &  9 & MYSO \\
K3-50 4 &      1.25 &     11.20 &    100.6 &      8.0 &  55.6 & - & 100.6 &   8.0 & - &  12.0 &  6 & MYSO \\
K3-50 5 &      5.23 &     10.84 &     52.8 &      8.0 &  42.4 & - &  82.2 &   8.0 & - &  24.0 &  8 & MYSO \\
K3-50 6 &      5.47 &    457.90 &     79.5 &     48.0 &  12.6 & - & 106.0 &   8.0 & - &  64.0 & 14 & MYSO \\
K3-50 7 &      2.53 &     13.30 &     53.0 &      8.0 &  26.5 & - &  75.5 &   2.0 & - &  16.0 & 16 & pMYSO$\dagger$\\
K3-50 8 &      6.24 &      7.70 &     26.5 &      4.0 &   8.4 & - &  56.2 &   4.0 & - &  16.0 & 16 & $\dagger$     \\
K3-50 C1 &    159.90 &    300.82 &    132.5 &     24.0 & 106.0 & - & 132.5 &  24.0 & - &  24.0 & 14 & MYSO \\
K3-50 C2 &     89.60 &    460.33 &    212.0 &     32.0 & 111.3 & - & 262.3 &  24.0 & - & 128.0 &  8 & MYSO 
\enddata
\tablecomments{\footnotesize  A ``MYSO'' in the right column denotes a MYSO candidate.}
\tablenotetext{\dagger}{\footnotesize The SED fits do not go through the 20\,$\mu$m data points. Therefore the fits are all under-fitting the data (and therefore underestimating the luminosities).}
\end{deluxetable*}

While the flux error in the flux calibration factor (Jy/ADU) of the FORCAST data is relatively small ($\le$15\%), the backgrounds around sources can be quite large and variable (i.e., not flat under the source), the fluxes obtained through background subtraction can carry a larger uncertainty.  This is true for the Spitzer and Herschel data as well. Since the upper limit uncertainty on the flux cannot be significantly larger than the background amount we subtracted, we set the upper error bar as the background flux value. The lower error bar values for all sources come from the average total photometric errors at each wavelength which are estimated to be 20\%, 15\% and 10\% for 4.5, 20 and 37\,$\mu$m bands, respectively. For K3-50 C1 and C2 only, the 25\,$\mu$m total photometric error is believed to be 25\% (see Section~\ref{sec:obs}). For most sources the Herschel data are used as upper limits since the resolutions in the data make it difficult to separate out flux from individual sources from other nearby sources or larger-scale environmental emission. However, there are some sources with sufficient isolation in the Herschel 70 and 160\,$\mu$m images that accurate background subtracted fluxes could be obtained (see Tables \ref{tb:PACS_DR7} and \ref{tb:PACS_K350}). For these sources we set the upper error bar as the background flux value, and for the lower error bar we estimate the photometric uncertainties to be 40\% and 30\% for 70 and 160\,$\mu$m bands, respectively.   

Once SEDs could be constructed from the photometric data (and their associated errors or limits), we utilized the ZT \citep{2011ApJ...733...55Z} MYSO SED model fitter as we did in \citetalias{2019ApJ...873...51L} in order to investigate the physical properties of individual sources. The fitter pursues a $\chi^2$-minimization to determine the best fit MYSO models, with each model fit providing a normalized minimum $\chi^2$ value (so called $\chi_{\rm nonlimit}^2$). To be consistent with the analysis of \citetalias{2019ApJ...873...51L}, we selected a group of models that show $\chi_{\rm nonlimit}^2$ values similar to the best fit model and distinguishable from the next group of models showing significantly larger $\chi_{\rm nonlimit}^2$ values (see further discussion in \citetalias{2019ApJ...873...51L}). 

Figure \ref{fig:DR7SED} plots the derived photometry points and shows the ZT MYSO SED model fits for each compact source in DR7, and in Figure \ref{fig:K350SED} the same for K3-50. The black lines represent the best model fit and gray lines are the rest in the group of best fit models. The number of the best fit models and the ranges of the derived parameters based on the models are listed in Table~\ref{tb:sed_DR7} for DR7 and Table~\ref{tb:sed_K350} for K3-50. One caveat is that the ZT models assume a single central stellar source, and given the relatively large distances to DR7 and K3-50, it is likely that at least some of our compact sources house multiple unresolved stellar components. It is for this reason that we do not tabulate or discuss the model parameters like disk size or accretion rate, and instead concentrate on the values for internal mass and source luminosity. An additional caution is that, since we do not have data points to constrain the depth of the silicate absorption feature at 10\,$\mu$m, the values of $A_V$ in the fits can sometimes vary widely. This is because the fitter can sometimes find equally good fits, for instance, with an edge-on disk and low extinction or with a face-on disk and high extinction.

The right columns of Tables~\ref{tb:sed_DR7} and \ref{tb:sed_K350} show the identification of the individual sources based on our criteria of MYSOs and possible MYSOs (``pMYSOs'') defined in \citetalias{2019ApJ...873...51L}. To summarize, the conditions for a source to be considered a MYSO is that it must 1) have an SED reasonably fit by the MYSO models, 2) have a M$_{\rm star}\ge8\,$M$_{\sun}$ for the best model fit model, and 3) have M$_{\rm star}\ge8\,$M$_{\sun}$ for the range of $M_{\rm star}$ of the group of best fit models. A pMYSO is any source that fulfills only the first two of these criteria. It appears that only one compact source, K3-50 7, fulfills the pMYSO criteria. Only one source in DR7 and two in K3-50 appear to be too low in mass to qualify as either a pMYSO or MYSO. 

Therefore, for K3-50, out of the ten SOFIA-FORCAST defined compact sources, eight satisfy our MYSO or pMYSO criteria ($80\%$). For DR7, we have four identified compact sources and three satisfy our criteria of housing a MYSO ($75\%$). The dearth of MYSOs in DR7 (3) is striking. Rather than comparing total MYSOs per G\ion{H}{2} region, a fairer comparison would be to compare the number of MYSOs per square parsec. We calculated the area containing the 3-$\sigma$ flux at 37\,$\mu$m for all sources we have so far studied. This includes DR7 (48\,pc$^2$) and K3-50 (32\,pc$^2$), as well as for G49.5-0.4 and G49.4-0.3 in W51A (107 and 69\,pc$^2$, respectively; \citetalias{2019ApJ...873...51L}), M17 (25\,pc$^2$; \citetalias{2020ApJ...888...98L}), and W49A (147\,pc$^2$; \citetalias{2021DeBuizer}). Interestingly, G49.5-0.4, M17, and K3-50 have similar values of MYSOs per square pc (0.29, 0.28, 0.25, respectively) even though they have quite different total MYSOs per region (31, 7, and 8, respectively). Both W49A (22 MYSOs) and G49.3-0.3 (10 MYSOs) have the same value of 0.15 MYSOs per square pc, a density half that of the previously mentioned regions, though for W49A this may be due to the large distance (11.1\,kpc) relative to the other regions ($2-8$\,kpc). DR7 indeed has the lowest value at 0.06 MYSOs/pc$^2$, substantially smaller than all other G\ion{H}{2} regions studied so far. 

DR7 is also an outlier among the other studied G\ion{H}{2} regions, in that its most massive MYSO, source 5, is only 16\,$M_{\sun}$ as determined via SED fitting. Since this source has radio continuum emission, we can confirm this mass using the radio continuum flux and source size measured by \citet{1994ApJS...91..659K}, and using the equations of \citetalias{2022DeBuizer}, we can derive a estimate of the log$N_{LyC}$ of 46.11, which is the equivalent of a B0.5 ZAMS star \citep{1973AJ.....78..929P} which has an estimated mass of 16.3\,$M_{\sun}$ \citep{2000AJ....119.1860B}. The mass of this most massive MYSO is more comparable to the the regions of Sgr D and W42 studied in \citetalias{2022DeBuizer} (16 and 32\,$M_{\sun}$, respectively), which are significant, but not giant, \ion{H}{2} regions. The most massive MYSO in K3-50 is 48\,$M_{\sun}$, is more comparable to the other G\ion{H}{2} regions we have studied. For comparison, the most massive MYSO in M17 was measured to be 64\,$M_{\sun}$, for G49.5-0.4 it was 96\,$M_{\sun}$, for G49.4-0.3 it was 64\,$M_{\sun}$, and for W49A it was 128\,$M_{\sun}$. Given the large distance to W49A, this large mass is likely due to an unresolved cluster of multi-MYSO system, but M17 and W51A are closer than DR7 and K3-50, and so distance is not the determinant factor.   

Sorted by distance, M17 (d$\sim1.98$kpc), G49.5-0.4 and G49.4-0.3 of W51A (d$\sim5.4$kpc), and W49 (d$\sim11.1$kpc) were found to have a percentage of MYSOs and pMYSOs to all compact sources of $44\%$, $90\%$, $70\%$ and $96\%$, respectively.  The main reason for the difference was suggested in \citetalias{2021DeBuizer} as being the result of the distance of the regions, i.e., M17 has a high detection rate of less-luminous low to intermediate mass YSOs due to its much closer proximity than, say, W49A. With K3-50 and DR7 lying at a distance between W51A and W49A, their percentages of MYSOs to total compact sources ($80\%$ and 75\%, respectively), are in rough agreement with that hypothesis, reinforcing the idea that distance does seem to play a role in how many lower-mass YSOs we can detect. We caution, however, that in the case of DR7 this percentage value should be viewed with less confidence due to very small number statistics. 

Of the three MYSO candidates in DR7, compact sources 5 and 6 are associated with cm radio continuum emission. The other compact source, source 4, therefore may be a MYSO in a stage of evolution prior to the onset of a UC\ion{H}{2} region. Source 4 lies at radii further out from the central revealed and ionizing star cluster than the ionized radio continuum region. This might be an area of swept-up material from the expanding ionization front of the revealed stellar cluster, which is now collapsing and forming a region where new star formation is taking place. For K3-50, only sources C1, C2, and source 8 appear to have radio continuum associated with their infrared-defined peaks. Therefore, the remaining sources (sources $1-7$) are also likely to be very young pre-ionizing MYSOs.  

\subsection{Physical Properties of Extended Sources: Kinematic Status and Global History}    
   
As we have done in our previous papers, here we attempt to investigate the evolutionary state and history of both DR7 and K3-50 by utilizing two different molecular clump evolutionary tracers, the luminosity-to-mass ratio ($L/M$) and unitless virial parameter ($\alpha_{vir}$) toward the sub-regions of each G\ion{H}{2} region. We assume these larger and extended radio continuum sub-regions are candidates for being star-forming clumps (rather than individual cores) housing embedded (proto)clusters of massive stars that are ionizing the extended \ion{H}{2} regions seen in radio continuum. Lower $\alpha_{vir}$ as well as $L/M$ values are assumed to demonstrate relatively younger star-forming clumps, and plotting the $\alpha_{vir}$ vs. $L/M$ parameters for the sub-regions within our previously studied G\ion{H}{2} regions yielded a relatively linear correlation. Ideally, we would wish to repeat this evolutionary analysis here for both DR7 and K3-50, but as we will discuss below, there exist insufficient data to do exactly the same analysis.  

For DR7, there are four extended sub-regions in the 20, 37, and 70\,$\mu$m maps that correlate with the major radio continuum sub-regions identified by \citet{1986ApJ...306..122O} in their 6.2\,cm maps. These sub-regions are radio sources A, B, C, and F (see Table \ref{tb:DR7_sources}). There are no definitive mid- or far-infrared peaks/sub-regions associated radio regions D, E, or G. For all extended sources, the value of the aperture to used for the photometry was determined by looking at the cm radio maps and finding an aperture that encompasses all of the cm emission from each source as well as the extended dust emission as seen in the 20, 37, and 70\,$\mu$m maps and determining the smallest aperture radius that would encompass each source at all of these wavelengths. Background subtraction was performed by the same methods as described in \citetalias{2019ApJ...873...51L}. These radii were then used for all photometry performed on data at SOFIA and Spitzer wavelengths (i.e. $R_{int}$ in Tables 8 and 10). Descriptions of how we determine the apertures for each source in the Herschel (and Spitzer) data are detailed in Appendix B. From this infrared photometry, we were able to derive the luminosity and mass values of each sub-region (as will be discussed in more detail below) necessary for the $L/M$ analysis. To perform the virial analysis of the identified sub-regions within DR7, we require molecular line maps. DR7 lies in projection within the Cygnus X region, and there are many molecular line maps of this part of the sky (e.g., $^{13}$CO ($J=2-1$) KOSMA maps of \citealt{2006A&A...458..855S}; $^{13}$CO ($J=1-0$), CS ($J=2-1$), and N$_2$H$^+$ ($J=1-0$) FCRAO maps of \citealt{2010A&A...520A..49S}; and JCMT $^{12}$CO ($J=3-2$) maps of \citealt{2012A&A...541A..79G}). However, since DR7 lies at a distance much farther than the rest of the Cygnus X sources, its molecular lines are shifted to a very different velocity range, and we can find no molecular maps that cover DR7 both spatially and in velocity range. 

For K3-50, we distinguish five sub-regions in the infrared maps that correlate to the major radio sub-regions A, B, C, D, and G, as identified in the 2\,cm radio continuum maps of \citet{1994ApJ...428..670D}. As described for DR7 above, we measured background-subtracted photometric values for all sub-regions in all wavelengths covered by Spitzer-IRAC, SOFIA, and Herschel. The radii employed for aperture photometry and the derived fluxes are given in Tables 9 and 11. Like DR7, we were not able to find CO maps of K3-50 with adequate spatial resolution for the purposes of our virial analyses. However, we were able to find HCO$^+$ ($J=4-3$) data from the James Clerk Maxwell Telescope (JCMT) data archive, and these observations were made with the Heterodyne Array Receiver Program \citep[HARP;][]{2009MNRAS.399.1026B} instrument, which has adequate angular resolution ($\sim$14$\arcsec$) to allow us to use it for the virial analyses. The map, however, covers two sub-regions of K3-50 only (A and B).

\subsubsection{Methodology and data used}

The derivations of mass ($M$) and bolometric luminosity ($L$) of each sub-region are based on the corresponding SED fitting methods described in \citet{2016ApJ...829L..19L} and \citetalias{2019ApJ...873...51L} and assume that each of these extended sub-regions is a molecular clump. In order to obtain each clump mass ($M$), we adopt the high-resolution graybody fitting method of \citet{2016ApJ...829L..19L}, whereby we first derive the low-resolution ($\sim$36$\arcsec$) temperature ($T$) map from the convolved $160 - 500\,\mu$m Herschel images (i.e., across cold dust temperature component). These low-resolution $T$ maps are then utilized as the templates to be applied to the higher resolution ($\sim$14$\arcsec$) dust emission maps from JCMT-SCUBA2 850$\mu$m archival data to obtain the final graybody fitted mass surface density ($\Sigma$) maps of DR7 and K3-50. We then estimate the mass of each sub-region by utilizing the $\Sigma$ maps with the distance to each source (i.e., 7.30~kpc for DR7; 7.64~kpc for K3-50). As in \citetalias{2019ApJ...873...51L}, the bolometric luminosities of the sub-regions are calculated based on the two temperature graybody fit which uses all wavelengths of Spitzer-IRAC, SOFIA, and Herschel. $L/M$ is believed to trace molecular clump evolution because, as more dust and gas from the star-forming molecular clump get incorporated into stars, the clump mass $M$ decreases (and thus $L/M$ increases). Similarly, as the molecular clump creates more stars, the infrared-derived $L$ will increase due to the heating caused by these additional new stars and the additional internal energetic stellar feedback they create (again, causing $L/M$ to increase).

To calculate the virial parameters ($\alpha_{\rm vir}$) for K3-50 A and B, we utilize the derived masses, distance estimates, and the full width half maximum (FWHM) of the HCO$^+$ data in km s$^{-1}$ (see Equation 2 of \citetalias{2019ApJ...873...51L}). Note that in all of our previous papers we have used $^{13}$CO ($J=1-0$) data for for the virial analyses because it is a good tracer for the kinematics of the dense ambient medium \citep[n$_{H_2}$$\lesssim$10$^4$ cm$^{-3}$;][]{2019PASJ...71S...3N}. However, HCO$^+$ ($J=4-3$) traces denser hydrogen molecules \cite[n$_{H_2}$$\lesssim$10$^5$ cm$^{-3}$;][]{2011A&A...525A.107R}, and possibly could be tracing the denser molecular core components within the our sub-regions (which we believe to be molecular clumps) rather than the sub-region as a whole. However, even though the derived $\alpha_{\rm vir}$ values of K3-50 A and K3-50 B utilize HCO$^+$ ($J=4-3$) line emission, we do not expect larger uncertainties when compared to the $\alpha_{\rm vir}$ values in Paper I, II, and III. This is because a recent study toward a Galactic star forming cloud (DR21) shows the velocity width of HCO$^+$ ($J=1-0$) and $^{13}$CO ($J=1-0$) are consistent \citep{bonne}. Smaller structures (molecular cores) were also previously studied by \citet{2004ApJ...612..946G} where they showed that the HCO$^+$ ($J=4-3$) and $^{13}$CO ($J=1-0$) velocity widths are almost identical. The virial parameter is believed to trace molecular clump evolution because its value is proportional to the square of the measured dispersion of the molecular line being studied, and such lines broaden as a clump evolves and more turbulent energy is injected into the clump by star formation processes. Therefore, larger measured virial parameters indicate more evolved molecular clumps.

\begin{deluxetable*}{rccccc}
\tabletypesize{\small}
\tablecolumns{6}
\tablewidth{0pt}
\tablecaption{Derived Parameters of Extended Sources in DR7}\label{tb:esDR7}
\tablehead{\colhead{   Source   } &
           \colhead{  $M$   } &
           \colhead{ $L$ } &
           \colhead{  $T_{\rm cold}$  } &
           \colhead{  $T_{\rm warm}$  } &
           \colhead{  $L/M$  }\\
	   \colhead{        } &
	   \colhead{ ($M_{\sun}$) } &
	   \colhead{ ($\times 10^4 L_{\sun}$) } &
	   \colhead{ (K) } &
	   \colhead{ (K) } &
       \colhead{  $L_{\sun}/M_{\sun}$  }
}
\startdata
    DR7 A &    110.9 &     4.85 &   65.3 &  256.7 &  218.5 \\
    DR7 B &     98.7 &     4.27 &   62.6 &  273.6 &  216.4 \\
    DR7 C &    318.8 &     5.68 &   55.1 &  285.7 &   89.2 \\
    DR7 F &     62.6 &     2.50 &   50.4 &  274.5 &  199.8 
\enddata
\end{deluxetable*} 

\begin{deluxetable*}{rccccccc}
\tabletypesize{\small}
\tablecolumns{8}
\tablewidth{0pt}
\tablecaption{Derived Parameters of Extended Sources in K3-50}\label{tb:esK3}
\tablehead{\colhead{   Source   }                                              &
           \colhead{  $M_{\rm vir}$   } &
           \colhead{  $M$   } &
           \colhead{ $L$ } &
           \colhead{  $T_{\rm cold}$  } &
           \colhead{  $T_{\rm warm}$  } &
           \colhead{  $L/M$  } &
           \colhead{ $\alpha_{\rm vir}$ }\\
	   \colhead{        } &
	   \colhead{ ($M_{\sun}$) } &
	   \colhead{ ($M_{\sun}$) } &
	   \colhead{ ($\times 10^4 L_{\sun}$) } &
	   \colhead{ (K) } &
	   \colhead{ (K) } &
       \colhead{  $L_{\sun}/M_{\sun}$  } &
       \colhead{     }
}
\startdata
    K3-50 A &   1993.6 &    845.2 &       144 &     69.0 & 212.4 &  850.7 &   2.36\\
    K3-50 B &   3037.9 &    659.4 &      32.3 &     68.8 & 284.1 &  244.6 &   4.61\\
    K3-50 C &  \nodata &    904.6 &      36.2 &     60.2 & 276.0 &  200.3 &\nodata\\
    K3-50 D &  \nodata &    152.4 &      24.7 &     69.7 & 266.6 &  811.3 &\nodata\\
    K3-50 G &  \nodata &     27.7 &      2.33 &     44.8 & 315.1 &  421.2 &\nodata
\enddata
\end{deluxetable*} 
    
\begin{figure}
\center
\begin{tabular}[b]{c@{\hspace{-0.0in}}c}
\includegraphics[width=3.2in]{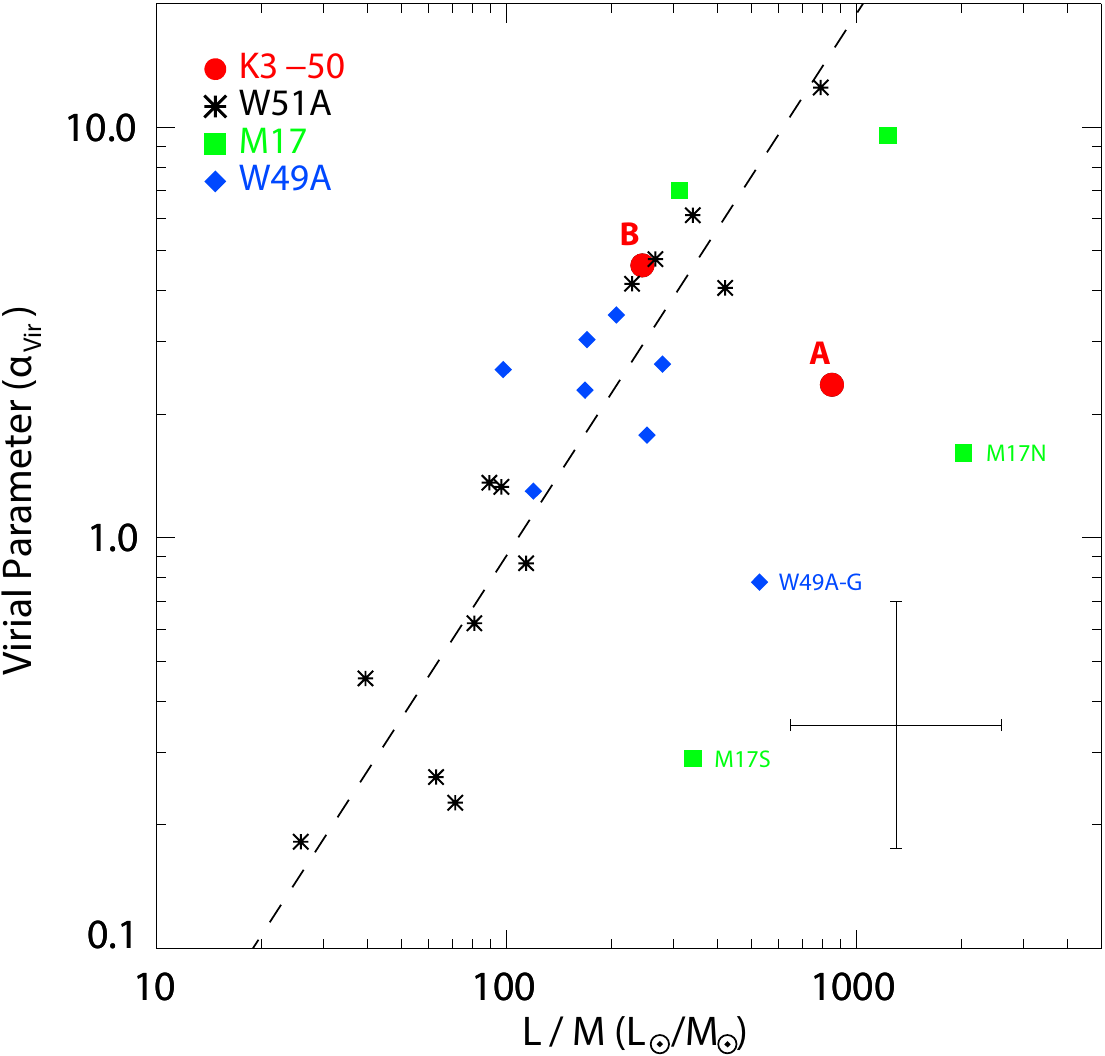}\\
\end{tabular}
\caption{\footnotesize Virial parameter ($\alpha_{\rm vir}$) vs. $L/M$ of all infrared sub-regions in all G\ion{H}{2} regions studied so far. Black asterisks are values for the sub-regions in W51A (i.e., both the G49.5-0.4 and G49.4-0.3 G\ion{H}{2} regions), and the  dashed line indicates the best line fit to the W51A data ($\alpha\sim$1.28 in log-space). Green squares show the sub-regions of M17, and blue diamonds show the data for the sub-regions in W49A. The red dots are the new data from this work for K3-50, with the sub-region A and B labels shown. While K3-50 B appears to align with the data trend seen in W51A, K3-50 A appears to have an inconsistently high $L/M$ value that places it in a part of the plot with the other outliers M17N, M17S, and source G in W49A. These regions are believed to have high $L/M$ values due to contamination by external heating/ionization. The error bar at the bottom left shows the typical uncertainty (a factor of $\sim$2) on both $L/M$ and $\alpha_{\rm vir}$.} \label{fig:evo}
\end{figure}

\subsubsection{Evolutionary Analyses: K3-50}

Since the available molecular maps only cover the A and B sub-regions of K3-50, they are the only two sources for which we could derive virial parameters. These values are given in Table \ref{tb:esK3}, and show that source A ($\alpha_{\rm vir}$ = 2.36) has a lower measured virial parameter than B ($\alpha_{\rm vir}$ = 4.61). Therefore, our virial analysis indicates that source A is younger than source B.  However, in contradiction to the virial analysis, the measured $L/M$ value for source B (245 $L_{\sun}/M_{\sun}$) is much smaller than that of source A (851 $L_{\sun}/M_{\sun}$), which would indicate that source B is more youthful than source A. 
 
 To determine which evolutionary analysis is correct, we look to previous studies for answers. \citet{2010ApJ...714.1015S} state that the source A is likely younger than B due to its association with intense $^{12}$CO ($J=1-0$) emission and high measured infrared excess. Additionally, \citet{1996ApJ...460..744H} claim that source A is less evolved than B due to its larger \ion{H}{2} region size. This is consistent with our discussion in Section \ref{sec:K350}, where we state that, given the large physical size of the radio region associated with source B at our adopted distance, it does not qualify as a youthful compact \ion{H}{2} region, and must be an older \ion{H}{2} region. These arguments would seem to indicate that our virial analysis is more accurate than our $L/M$ analysis for sources A and B, and that source A is indeed more youthful than B.
 
We have seen cases in our previous papers in this survey where the $L/M$ values for some sources appear to be larger than they should be when compared to the evolutionary states that were derived based upon the virial parameters. To demonstrate this, we show sub-regions A and B on a plot of $\alpha_{\rm vir}$ vs. $L/M$ in Figure \ref{fig:evo}. The plot also shows data for all of the sub-regions measured so far across all G\ion{H}{2} regions studied as a part of this survey. The dashed line shown is a fit to the W51A sub-region data, and shows that $L/M$ and $\alpha_{\rm vir}$ are well-correlated. Furthermore, we see that K3-50 B and the data for most of the sub-regions measured so far across all G\ion{H}{2} regions are in reasonable agreement with the W51A trend. However, K3-50 A, along with M17N, M17S, and W49A G, do not fit the trend well. For M17N, M17S, and W49A G it was argued (in \citetalias{2020ApJ...888...98L} and \citetalias{2021DeBuizer}) that the large $L/M$ value was likely due to external heating by a nearby older generation of stars. Since the data point for K3-50 A resides in the same location of the plot as these other sources, this may be the same reason why K3-50 A has such a large measured $L/M$ value as well. However, while there is less obvious sources of external heating in the case of K3-50 A than in the cases of the other contaminated sub-regions mentioned, some evidence does exist. \citet{1977MNRAS.179..255W} claim that only 1/5th of the total infrared luminosity of K3-50 A can be attributed to the the emission from within the ionized region, the rest comes from outside it. \citet{1980ApJ...236..465I} find a nearby ($\sim$4$\arcsec$ away) reddened star called A2, that may be a source of additionally heating and/or ionization. However, while additional heating due to the presence of further nearby stars may be possible, \citet{2010ApJ...714.1015S} found that determining which stars are associated with the K3-50 region was difficult because of significant pollution from luminous field stars. 

Apart from source A, the results from the $L/M$ analysis of all of the sub-regions within K3-50 do appear to correspond well with the relative ages of the sources speculated via other means and investigators. Our $L/M$ values indicate that source D (811 $L_{\sun}/M_{\sun}$) is likely the most evolved, and source C (200 $L_{\sun}/M_{\sun}$) should be the youngest. Previous observations by \citet{1996ApJ...460..744H} and \citet{2010ApJ...714.1015S} appear to agree with this analysis. \citet{2010ApJ...714.1015S} further speculate that since sub-region B has more intense near-infrared and radio emission compared to D that it should be younger, which also consistent with the analysis by \citet{1996ApJ...460..744H} and our $L/M$ analysis. In summary, based upon our analyses and those of previous studies it is likely that the order of relative ages of the sub-regions of K3-50, from youngest to oldest, are: C, A, B, G, and D. 

With this information concerning the relative ages of the sub-regions, we can speculate on the hypotheses of the origin of K3-50 as a whole. There are two main scenarios discussed in the literature for the origin of K3-50. The first is from \citet{1980ApJ...236..465I}, who observe a large bubble of \ion{H}{1} emission going through K3-50 with a center located somewhere between 12$\arcmin$ and 14$\arcmin$ to the southwest of K3-50. They speculate that the bubble was created by strong stellar winds from an older cluster of OB stars. However, observations by \citet{1998ApJS..115..271R} found only a few B-type stars situated (by projection) inside the alleged \ion{H}{1} bubble and the distances to these star are not known with enough certainty to say whether or not they are related to the \ion{H}{1} emission. \citet{2010ApJ...714.1015S} suggest that the \ion{H}{1} shell may have been created by the Wolf-Rayet star WR 131 instead (which is situated to the southwest of K3-50 but not at the center of the \ion{H}{1} shell). They claim that the location of source D closer to the shell center and being older than source C (which is farther from the center) argues favorably for the shell expansion scenario, since the expansion would have triggered the formation of D first. However, if this were the case, one would expect source B to be younger than source A as well (since sub-region B is farther from the alleged bubble center than sub-region A), and this is not consistent with the relative ages argued by us or \citet{1996ApJ...460..744H}, nor even \citet{2010ApJ...714.1015S}.   

The second scenario for the origin of K3-50 is from \citet{1996ApJ...460..744H} who claim that the sources C, B, A, and D are distributed in an arc, and if fit with an ellipse, the center would be 110$\arcsec$ east of K3-50 A. They claim it may be that a supernova event at this location could have triggered all the star formation in K3-50. However, there is no further support for the supernova scenario as there is no known supernova remnant at the ellipse center. Furthermore, if the speculation is that all of the sub-regions of K3-50 can be fit by an ellipse because they were triggered to form at the same time by the supernova shock, then all of the sub-regions would have very similar evolutionary states. However, our measured $L/M$ values for K3-50 range from 200 to 851 $L_{\sun}/M_{\sun}$ (or 200 to 811 $L_{\sun}/M_{\sun}$ if we disregard source A), which is a very large range of $L/M$ values, and  comparable to the large range seen in W51A ($26 < L/M < 790 L_{\sun}/M_{\sun}$; see Figure \ref{fig:evo}), which is believed to have undergone multiple separate star formation events separated widely in time \citepalias{2019ApJ...873...51L}. 

Therefore, the relative evolutionary states of the sub-regions of K3-50 do not seem to support either previously suggested global formation scenario. Rather than being due to wide scale or global triggers, we instead suggest that the present appearance of K3-50 is the product of multiple star forming events separated more widely in time. 

\subsubsection{Evolutionary Analyses: DR7}

Since we do not have any molecular data maps covering the velocity range of DR7, we cannot derive the viral parameters for any of the sub-regions within it. Therefore, we can only use the $L/M$ results to infer something about the evolution of DR7. Furthermore, unlike K3-50, there is no discussion in the literature about the evolution or origin of DR7, so we will speculate on its nature here.

DR7 appears to be a cavity that has been carved by a previous generation of massive stars. At the end of Section \ref{sec:DR7}, we speculate that the present star formation in DR7 is the result of the expansion of a bubble around the Cl09 star cluster as it expands into a molecular cloud to the northwest. Looking at the location of the sub-regions in DR7, we see that A, B, and F all lie on the inner edge of the cavity, and thus are all approximately the same distance from the central cluster of stars. If star formation in all these sub-regions was triggered by the material swept up by the radiation pressure from the Cl09 cluster they should all have similar evolutionary states. Consistent with this, the $L/M$ ratios of these sub-regions have a small range of values between 200 and 219 $L_{\sun}/M_{\sun}$. The sub-region C has a $L/M$ ratio of 89 $L_{\sun}/M_{\sun}$ which is very different from the others but it is located (in projection) further from the central cluster of stars. Therefore, it may be that the C sub-region is younger than the other sub-regions because star formation was triggered on the inner cavity walls first and is now proceeding to locations further away, consistent with our hypothesis. 

Furthermore, the spread of $L/M$ values for the sub-regions inside DR7 is relatively modest. The $L/M$ values for DR7 only range from 89 $L_{\sun}/M_{\sun}$ to 292 $L_{\sun}/M_{\sun}$, which is more comparable to the $L/M$ range seen in W49A \citepalias{2021DeBuizer}, which has measured values from 83 to 281 $L_{\sun}/M_{\sun}$ (disregarding source G which has a $L/M=539$ $L_{\sun}/M_{\sun}$ but is also thought to be externally contaminated). For W49A it is argued that this small spread in evolutionary state of its sub-regions supports the hypothesis that a coeval starburst event was responsible for most or all of the star formation presently being observed. Similarly, therefore, the small range in $L/M$ for DR7 may be indicating that the star formation is more or less coeval in this region as well.

\subsection{Are DR7 and K3-50 Genuine Giant H II Regions?} 

As discussed in \citetalias{2022DeBuizer}, defining a bona-fide G\ion{H}{2} region by the cut-off criterion of $N_{LyC} = 10^{50}$\,photons/s, is somewhat arbitrary given that the entire population of ionized regions of the Galaxy exist on a continuum of $N_{LyC}$ values. However, the moniker of ``giant'' \ion{H}{2} region is meant to be a designation for those regions that house the largest clusters of young O stars and represent the main locations of massive star formation in the Galaxy. As stated in \citet{2004MNRAS.355..899C}, this value is the equivalent to the photon rate of more than 10 O7V stars, and \citet{1970IAUS...38..107M} define this cut-off as regions larger than 4 times that of Orion, and as such should be easily discernible in the spiral structures of external galaxies. 

In \citetalias{2022DeBuizer} we studied Sgr D and W42, two sources below the $N_{LyC} = 10^{50}$\,photons/s cut-off but with values close to that of Orion (i.e., 49.47 photon/s; \citealt{2001ApJ...555..613I}), and found that they are each ionized predominantly by a single O star (as Orion is predominantly ionized by the O6V star, $\theta^1$ Orionis C1; \citealt{2014AstBu..69...46B}). On the other hand, regions like NGC\,3603 with $N_{LyC} > 10^{51}$\,photons/s must be powered by many O stars, since no single O star can have that photon rate. Therefore, the value of $N_{LyC}$ is useful to some degree in helping distinguish between \ion{H}{2} regions powered by single O star and those powered by large clusters of O stars. However, near this cut-off value of $N_{LyC} = 10^{50}$\,photons/s it is possible to have a region that is either powered predominantly by a single very massive O star (e.g., an O4V star) or by a cluster of slightly more modest O stars (e.g., 10 O7V stars). By looking at the infrared properties of the regions we have studied so far, in \citetalias{2022DeBuizer} we discuss how we might be able to use the Lyman continuum photon rate in conjunction with other observational indicators in the infrared to distinguish between regions that are the birthplaces massive clusters (i.e., G\ion{H}{2} regions) versus those that simply house one (or a few) O stars. Both DR7 and K3-50 have $N_{LyC}$ values just above the cut-off criterion, with 1-$\sigma$ errors that dip below the cut-off criterion, placing doubt on their classification as G\ion{H}{2} regions based upon their $N_{LyC}$ values alone. 

In \citetalias{2022DeBuizer}, we discussed four observational characteristics in the thermal infrared that appeared to differentiate bona-fide G\ion{H}{2} regions from large G\ion{H}{2} regions: 1) the number of compact infrared sources (as seen by SOFIA), 2) the number of sub-regions, 3) the percentage of flux in the brightest peak at 37\,$\mu$m, and 4) the mass of the most massive MYSO (from our SED modeling).  For all \ion{H}{2} regions studied so far in our series of papers, including DR7 and K3-50, we present the measured values for each of these indicators in Table \ref{tb:properties} listed in order (top-to-bottom) by number of compact sources. We see that K3-50 has the same number of compact sources (10) as the G\ion{H}{2} region W51A: G49.4-0.3, while DR7 has only 4 compact sources, making it more comparable to the (non-giant) \ion{H}{2} regions Sgr~D and W42. Likewise the number of radio sub-regions in K3-50 (5) is comparable to the G\ion{H}{2} regions W51A: G49.4-0.3 (5) and M17 (4), whereas DR7 has one main radio sub-region. We already discussed in the previous section that the mass of the most massive MYSO in DR7 (16\,$M_{\sun}$) is comparable to the (non-giant) Sgr\,D \ion{H}{2} region, whereas K3-50 has a most massive MYSO (48\,$M_{\sun}$) below, but comparable to M17 and W51A: G49.4-0.3 (both of which are 64\,$M_{\sun}$). Contrary to this trend however, it appears that K3-50A dominates the emission in the region, making up 59\% of the total flux of the entirety of K3-50, similar to what we have seen for the (non-giant) \ion{H}{2} region W42. Whereas, source 5 in DR7, though the brightest source in the region, only accounts for 15\% of the entire 37\,$\mu$m flux of DR7, more akin to the G\ion{H}{2} regions W51A: G49.4-0.3 and W51A: G49.5-0.4. Therefore, three of the four indicators that were suggested in \citetalias{2022DeBuizer} would seem to suggest that K3-50 is likely to be a genuine G\ion{H}{2}, and suggest DR7 is not. The only suggested indicator that runs contrary to the others for DR7 and K3-50 is the percentage of flux in the brightest peak. We therefore tentatively classify K3-50 as a G\ion{H}{2} region and DR7 as a \ion{H}{2} region.

\begin{deluxetable*}{lcccccc}
\tabletypesize{\small}
\tablecolumns{12}
\tablewidth{0pt}
\tablecaption{Infrared Observational Properties of All Surveyed \ion{H}{2} Regions to Date }\label{tb:properties}
\tablehead{\colhead{Region}                                              &
           \colhead{ No. Compact Sources  } &
           \colhead{ No. Sub-Regions  } &
           \colhead{ \%Flux in Peak} &
           \colhead{  Highest Mass YSO } &
           \colhead{ Type }
}
\startdata
W51A: G49.5-0.4  &      37 &      10 &      20 &       96 & G\ion{H}{2}\\
W49A             &      24 &      15 &      25 &      128 &G\ion{H}{2}\\
M17              &      16 &       4 &       5 &       64 & G\ion{H}{2}\\
W51A: G49.4-0.3  &      10 &       5 &      15 &       64 & G\ion{H}{2}\\
K3-50            &      10 &       5 &      59 &       48 & G\ion{H}{2}?\\
DR7              &      4  &       1 &      15 &       16 & \ion{H}{2}?\\
Sgr D            &      3  &       3 &      85 &       16 & \ion{H}{2}\\
W42              &      2  &       1 &      50 &       32 & \ion{H}{2}
\enddata
\end{deluxetable*}

However, there may be an additional complexity that could make comparing all G\ion{H}{2} regions using the above assessments more uncertain and show that the methods we have used so far are not perfect. There is a morphological similarity between DR7 to M17, and these sources lie in contrast to the other regions we have studied. Both DR7 and M17, as seen in their dust emission in the infrared, are predominantly cavity structures, where a revealed stellar cluster from a previous epoch of star formation may be responsible for a significant amount (if not the vast majority) of the ionization and dust heating of the entire cavity. The Lyman continuum emission from recent and presently-forming massive stars (and small massive star clusters) may play only a minor role in the overall Lyman continuum flux of the whole region. This stands in contrast to W51A:G49.5-0.4, W51A:G49.4-0.3, W49A, and K3-50, which all seem to be collections of dusty, ionized sub-regions confined to separate large molecular clumps containing recent and presently-forming massive stars (and small massive star clusters), whose combined ionization supplies the majority of the overall Lyman continuum flux of the region. In these regions, the Lyman continuum contribution of previous epochs of star formation seem to play a smaller role. Therefore, in M17 and DR7, it may be that the extremely energetic burst of previous star formation may have also caused significant feedback in the form of radiation pressure, clearing material in the immediate vicinity of the cluster and stifling any immediately subsequent generation of stars from forming. However, over time such radiation pressure snowplowed the material in all directions until the material in the cavity walls was dense enough (or the cavity impinged on an already present molecular clump) such that it collapsed locally into stars (e.g.,  such stellar feedback processes have been observed and studied in the G\ion{H}{2} region RCW49 by \citealt{2021ApJ...914..117T}). In both M17 and DR7 the MYSOs and compact sources we are finding are indeed concentrated in or near their cavity walls. In this way, regions like M17 and DR7 may have ``missing'' generations of star formation, unlike regions like W51A: G49.5-0.4, and thus have far fewer MYSOs in total. The dearth of MYSOs may be an indication that DR7 and M17 are, overall, more evolved G\ion{H}{2} regions compared to the others, or are close to being in a state between major star formation events. In addition to the lack of compact sources, being that the majority of the infrared and radio emission comes from ridges delineating the cavity walls, regions like M17 and DR7 have far fewer sub-regions of radio emission. Consequently, looking at the number of compact infrared sources (i.e. YSOs) and/or number of radio sub-regions may not be useful in determining if a region is a G\ion{H}{2} region if it has a cavity morphology. Looking at the available Spitzer images of the top ten most powerful G\ion{H}{2} regions (as given by their $N_{LyC}$ values in Paper IV), we see that three of these regions, M17, NGC 3603, and RCW49, appear to be large-scale cavity structures, indicating that this morphological class of G\ion{H}{2} regions may be a significant portion of the overall population of G\ion{H}{2} regions.

\section{Summary}\label{sec:sum}

In this, our fifth paper from our mid-infrared imaging survey of Milky Way Giant \ion{H}{2} regions, we obtained SOFIA-FORCAST 20/25 and 37\,$\mu$m maps toward the sources DR7 and K3-50, covering the most intense infrared-emitting areas of each region at $\sim$3$\arcsec$ spatial resolution. The 37\,$\mu$m images are the highest spatial resolution infrared observations of the entirety of both regions yet obtained at wavelengths beyond 25\,$\mu$m. We compared these SOFIA-FORCAST images with previous multi-wavelength observations from the near-infrared to radio wavelengths in order to inspect the morphological and physical properties of the compact and extended sources within in the DR7 and K3-50 complexes. We itemize below our main conclusions from this study.

Of the seven main radio peaks identified in DR7, we see clear 20 and 37\,$\mu$m infrared dust continuum associated with only A, B, C, and F. We see a ridge of infrared emission at the location of E, but there is no definitive peak, and sources D and G correspond to deficits in the SOFIA infrared emission maps. The brightest source on the SOFIA field at 20 and 37\,$\mu$m is the point-like infrared source 5, which is associated with the compact radio source G79.320+1.313. Overall, the extended dust emission seen in the SOFIA data is arc-shaped, with an apex to the northwest. The Spitzer data show this to be the brighter portion of the rim of a complete bubble that was likely created by the evolved Cl09 stellar cluster discovered by \citet{2002A&A...392..869L}. We speculate that the present star formation in DR7 was triggered by the expansion of this bubble into a molecular cloud to the northwest of the stellar cluster. 

In K3-50, we detect extended infrared emission in the direction of the previously discovered N-S outflow of source A at both 20 and 37\,$\mu$m, and the southern lobe of the outflow is brighter at both wavelengths, consistent with that side being the blue-shifted outflow lobe. At our adopted distance to K3-50 of 7.64\,kpc, both the B and D sources are too big to be considered compact \ion{H}{2} regions and instead are likely more evolved and expanding \ion{H}{2} regions. In both sources, the shell of gas and dust seen in the radio and infrared are likely the remains of a cavity created by a massive O star, displaying brighter 20\,$\mu$m emission from the hotter dust closer to the interior, while the 37 and 70\,$\mu$m emission are a better tracer of the cooler dust in the outer parts of the \ion{H}{2} region shell. While source C2 is more prominent between 2 and 25\,$\mu$m, C1 is brighter at 37\,$\mu$m and longer. We speculate the infrared extension of emission we are seeing in the SOFIA and Hershel data for C1 may be the dust cavity walls or the dust in the outflow from the stellar source at the heart of C1. We also identify a new region, which we label source G, and identify a nearby NIR stellar source that may be heating and ionizing it.

Of the ten compact or extended sources identified in the infrared for DR7, five are newly identified for the first time here. For K3-50, we identified 14 compact or extended sources, of which eight are newly identified. While three of the previously identified radio peaks in DR7 have no associated infrared peaks, we detect all previously identified radio continuum sources in K3-50 in the SOFIA maps. From our SED modeling of the near- to far-infrared emission of just the compact infrared sources identified in our SOFIA data, eight out of ten (80\%) satisfy our criteria for housing a MYSO (or pMYSO) in K3-50, and three out of four (75\%) satisfy our criteria in DR7. One of the three MYSOs in DR7 is not associated with radio continuum emission, and in K3-50, seven of the ten sources do not have associated radio continuum. We speculate that these sources are extremely young MYSOs at an early evolutionary stage prior to the onset of ionized emission.

DR7 has by far the lowest total number of MYSOs (3) and the lowest MYSO density (0.06 MYSOs/pc$^2$) than all other G\ion{H}{2} regions studied so far. DR7 is also an outlier among the other studied G\ion{H}{2} regions in that its most massive MYSO (source 5) is only 16\,$M_{\sun}$ as determined by our SED fitting. By comparison, K3-50 has a similar MYSO density (0.25 MYSOs/pc$^2$) to our previously studied G\ion{H}{2} regions W51A:G49.5-0.4 and M17 (0.29 and 0.28 MYSOs/pc$^2$, respectively), and the most massive MYSO in K3-50 (source 6) is 48\,$M_{\sun}$, again more comparable to the other G\ion{H}{2} regions we have studied.

In our analysis of the relative evolutionary states of the sub-regions within DR7, we find that the C sub-region is likely younger than the other sub-regions (A, B, and F), which all seem to be in a slightly older yet similar evolutionary state. Sub-regions A, B, and F are all on the edge of the large-scale arc of dust emission approximately equidistant from the Cl09 star cluster, and the C sub-region is further away from the cluster, which is consistent with our hypothesis that the present star formation in DR7 is the result of the expansion of a bubble around the Cl09 star cluster as it expands into a molecular cloud to the northwest. 

For K3-50, the results from our evolutionary analyses for the sources correspond well with the relative ages of the sources speculated via other means. Our $L/M$ values indicate that sub-region D is likely the most evolved, and source C should be the youngest. The results from our virial analysis indicate that sub-region A is likely very young as well, and younger than B. The large age spread found in the $L/M$ values for the sub-regions in K3-50 would indicate that the star-forming complex has undergone multiple star forming events separated widely in time. 

In \citetalias{2022DeBuizer} we suggested the use of several secondary indicators that might differentiate whether a region is a genuine G\ion{H}{2} region or not when the measured $N_{LyC}$ value has a large error and/or is near the qualifying cutoff. These indicators include the number of compact infrared sources present, number of sub-regions, the percentage of overall infrared flux from the region contained in the brightest source, and the mass of the highest mass MYSO. K3-50 has values more akin to genuine G\ion{H}{2} regions, whereas DR7 has values more like those of the non-G\ion{H}{2} regions studied in \citetalias{2022DeBuizer}. However, we are finding that population of G\ion{H}{2} regions may contain two distinctly different morphological types, those with distributed radio sub-regions and those with contiguous cavity structures, and their evolution (and thus observed properties) may differ significantly, suggesting the above indicators may be perhaps too simplistic.

\begin{acknowledgments}
The authors would like to thank the constructive input and advice of Nicole Karnath and Lars Bonne. This research is based on observations made with the NASA/DLR Stratospheric Observatory for Infrared Astronomy (SOFIA). SOFIA is jointly operated by the Universities Space Research Association, Inc. (USRA), under NASA contract NAS2-97001, and the Deutsches SOFIA Institut (DSI) under DLR contract 50 OK 0901 to the University of Stuttgart. This work is also based in part on archival data obtained with the Spitzer Space Telescope, which is operated by the Jet Propulsion Laboratory, California Institute of Technology under a contract with NASA. This work is also based in part on archival data obtained with Herschel, an European Space Agency (ESA) space observatory with science instruments provided by European-led Principal Investigator consortia and with important participation from NASA.  
\end{acknowledgments}

\vspace{5mm}
\facility{SOFIA(FORCAST)}

\clearpage

\appendix

\section{Data release}

The fits images used in this study are publicly available at: {\it https://dataverse.harvard.edu/dataverse/SOFIA-GHII}. 

The data include the SOFIA FORCAST 20 and 37\,$\mu$m final image mosaics and their exposure maps for all of DR7 and the central area of K3-50 (i.e. regions A, B, D, and G). Also included are the individual 25 and 37\,$\mu$m images of the K3-50 C region (i.e., sources C1 and C2).

\section{Additional Photometry of Sources in DR7 and K3-50}\label{sec:app}

As discussed in \S\ref{sec:data}, in addition to the fluxes derived from the SOFIA-FORCAST data, we used some additional photometry data in or SED analyses from both Spitzer-IRAC and Herschel-PACS.  

As we mentioned in \S\ref{sec:cps}, we performed optimal extraction photometry for the FORCAST 20 and 37\,$\mu$m images to define the location of all compact sources, and to determine the aperture radii to be used for photometry. Using these source locations, we employed the optimal extraction technique on the Spitzer-IRAC data for all sources to find the optimal aperture for each wavelength. As we have done for the FORCAST images, we estimated the background emission from the annuli that showed the least contamination from nearby sources, i.e. showing relatively flat radial intensity profile (\S\ref{sec:cps}). Table\,\ref{tb:IRAC_DR7} shows the photometry values we derive for all sources from the Spitzer-IRAC bands for DR7 and Table\,\ref{tb:IRAC_K350} shows the photometry values we derive for all sources from the Spitzer-IRAC bands for K3-50.

Table\,\ref{tb:PACS_DR7} shows the photometry result for the Herschel-PACS bands for DR7 and Table\,\ref{tb:PACS_K350} shows the results for K3-50. We attempted to use the optimal extraction technique for all sources to determine their aperture radii for photometry, however this was sometimes failed due to the ubiquity of extended emission from nearby sources that are overlapping the source being measured. For sources sufficiently isolated from contamination where we could perform an optimal extraction, we performed background subtraction as well. For the remaining sources, we used an aperture that best fit the largest size of the source at any wavelength. We believe that these aperture sizes are reasonable, especially since the data are only being used to provide upper limits to our SED model fits. 

\begin{deluxetable*}{lrrrrrrrrrrrr}
\tabletypesize{\scriptsize}
\tablecaption{Spitzer-IRAC Observational Parameters of Sources in DR7}
\tablehead{\colhead{  }&
           \multicolumn{3}{c}{${\rm 3.6\mu{m}}$}&
           \multicolumn{3}{c}{${\rm 4.5\mu{m}}$}&
           \multicolumn{3}{c}{${\rm 5.8\mu{m}}$}&
           \multicolumn{3}{c}{${\rm 8.0\mu{m}}$\tablenotemark{\tiny{a}}}\\
           \cmidrule(lr){2-4} \cmidrule(lr){5-7} \cmidrule(lr){8-10} \cmidrule(lr){11-13}\\
           \colhead{ Source }&
           \colhead{ $R_{\rm int}$ } &
           \colhead{ $F_{\rm int}$ } &
           \colhead{ $F_{\rm int-bg}$ } &
                      \colhead{ $R_{\rm int}$ } &
           \colhead{ $F_{\rm int}$ } &
           \colhead{ $F_{\rm int-bg}$ } &
                      \colhead{ $R_{\rm int}$ } &
           \colhead{ $F_{\rm int}$ } &
           \colhead{ $F_{\rm int-bg}$ } &
                      \colhead{ $R_{\rm int}$ } &
           \colhead{ $F_{\rm int}$ } &
           \colhead{ $F_{\rm int-bg}$ } \\
	   \colhead{  } &
	   \colhead{ ($\arcsec$) } &
	   \colhead{ (mJy) } &
	   \colhead{ (mJy) } &
	   \colhead{ ($\arcsec$) } &
	   \colhead{ (mJy) } &
	   \colhead{ (mJy) } &
	   \colhead{ ($\arcsec$) } &
	   \colhead{ (mJy) } &
	   \colhead{ (mJy) } &
	   \colhead{ ($\arcsec$) } &
	   \colhead{ (mJy) } &
	   \colhead{ (mJy) } \\
}
\startdata
\multicolumn{13}{c}{Compact Sources} \\
\hline
DR7 2  &	3 &	11.6	&5.99	&4   &	13.0	&   5.63	&5  &199	    &93.6	    &6	&642    	&284  \\
DR7 4  &	6 &	48.8	&24.2	&6   &	52.2	&   22.9	&6	&299    	&151    	&8	&1200   	&516  \\
DR7 5  &	6 &	83.7	&64.7	&6   &	119     &	99.5	&6	&499	    &345    	&\nodata	&\nodata	    &\nodata      \\
DR7 6  &	6 &	39.8	&13.6	&7   &	43.0	&   13.0	&7	&442     	&145    	&8	&1000   	&391   \\
\hline
\multicolumn{13}{c}{Extended Sources} \\
\hline
DR7 1  &    12&	80.5	&24.6	&12  &	77.7    &	17.1	&12	&728	    &213    	&13	&1980   	&827  \\
DR7 3  &	24&	589	    &293	&24  &	600     &	113  	&30	&1890   	&1780   	&30	&13800	    &4850     \\ 
DR7 A  &	14&	295  	&129	&14  &	344     &	146  	&14	&1310    	&566    	&14	&3140  	    &986  \\
DR7 B  &	14&	258	    &127	&14  &	323     &	187  	&14	&1740   	&1160   	&14	&4080   	&2550  \\
DR7 C  &	18&	510	    &2605	&18  &	524     &	257 	&18	&2930   	&1240   	&18	&6320	    &2580  \\
DR7 F &	10&	125	    &57.6	&10  &	121     &	63.8	&17	&1690   	&843    	&17	&3330   	&1570  \\
\enddata
\tablecomments{Entries with no data at 8\,$\mu$m are saturated in that band.}
\tablenotetext{a}{Entire field containing DR7 is contaminated with array artifacts at 8\,$\mu$m due to the saturated pixels from source 5 adding extra uncertainty to the 8\,$\mu$m photometry of all sources.}
\label{tb:IRAC_DR7}
\end{deluxetable*}
 
\begin{deluxetable*}{lrrrrrrrrrrrr}
\tabletypesize{\scriptsize}
\tablecaption{Spitzer-IRAC Observational Parameters of Sources in K3-50}
\tablehead{\colhead{  }&
           \multicolumn{3}{c}{${\rm 3.6\mu{m}}$}&
           \multicolumn{3}{c}{${\rm 4.5\mu{m}}$}&
           \multicolumn{3}{c}{${\rm 5.8\mu{m}}$\tablenotemark{\tiny{a}}}&
           \multicolumn{3}{c}{${\rm 8.0\mu{m}}$\tablenotemark{\tiny{a}}}\\
           \cmidrule(lr){2-4} \cmidrule(lr){5-7} \cmidrule(lr){8-10} \cmidrule(lr){11-13}\\
           \colhead{ Source }&
           \colhead{ $R_{\rm int}$ } &
           \colhead{ $F_{\rm int}$ } &
           \colhead{ $F_{\rm int-bg}$ } &
                      \colhead{ $R_{\rm int}$ } &
           \colhead{ $F_{\rm int}$ } &
           \colhead{ $F_{\rm int-bg}$ } &
                      \colhead{ $R_{\rm int}$ } &
           \colhead{ $F_{\rm int}$ } &
           \colhead{ $F_{\rm int-bg}$ } &
                      \colhead{ $R_{\rm int}$ } &
           \colhead{ $F_{\rm int}$ } &
           \colhead{ $F_{\rm int-bg}$ } \\
	   \colhead{  } &
	   \colhead{ ($\arcsec$) } &
	   \colhead{ (mJy) } &
	   \colhead{ (mJy) } &
	   \colhead{ ($\arcsec$) } &
	   \colhead{ (mJy) } &
	   \colhead{ (mJy) } &
	   \colhead{ ($\arcsec$) } &
	   \colhead{ (mJy) } &
	   \colhead{ (mJy) } &
	   \colhead{ ($\arcsec$) } &
	   \colhead{ (mJy) } &
	   \colhead{ (mJy) } \\
}
\startdata
\multicolumn{13}{c}{Compact Sources} \\
\hline
K3-50 1	&	5	&	22.5	&	12.6	&	5	&	34.1	&	23.4	&	5	&	157 	&	62.6	&	5	&	316		    &	108  \\
K3-50 2	&	7	&	56.9	&	33.9	&	7	&	77.0	&	47.6	&	7	&	361 	&	162 	&	7	&	765		    &	370 \\
K3-50 3	&	8	&	106		&	37.9	&	8	&	121 	&	56.2	&	5	&	334 	&	134 	&	8	&	1660		&	521   \\ 
K3-50 4	&	4	&	14.9	&	2.78	&	4	&	14.8	&	3.79	&	4	&	87.8 &	\nodata	&	4	&	200	    &	\nodata 			\\
K3-50 5	&	4	&	39.3	&	13.4	&	5	&	57.7	&	9.75	&	6	&	601 	&	114 	&	6	&	1180		&	\nodata			\\
K3-50 6	&	4	&	45.9	&	19.8	&	4	&	46.5	&	22.5	&	5	&	674 	&	241 	&	\nodata	&	sat			&	sat 		 \\
K3-50 7	&	4	&	37.8	&	6.30	&	4	&	38.3	&	5.28	&	4	&	232 	&	23.2	&	5	&	1290		&	\nodata    		\\
K3-50 8	&	4	&	36.1	&	7.70	&	4	&	39.6	&	9.65	&	6	&	415 	&	52.0	&	6	&	886		    &	149	 \\
K3-50 C1	&	4	&	8.57	&	\nodata	&	4	&	22.1	&	\nodata	&	4	&	130  &	\nodata	&	4	&	280	    &	\nodata  		\\
K3-50 C2	&	5	&	186		&	137 	&	6	&	355 	&	257 	&	\nodata	&	sat		&	sat		&	\nodata	&	sat			&	sat  		\\
\hline
\multicolumn{13}{c}{Extended Sources} \\
\hline
K3-50 A	&	\nodata	&	sat		&	sat		&	\nodata	&	sat		&	sat		&	\nodata	&	sat		&	sat		&	\nodata	&	sat			&	sat	 		\\
K3-50 B	&	36	&	2800	&	1940	&	36	&	2990	&	2380	&	36	&	20200	&	14900	&	\nodata	&	sat			&	sat	 		\\
K3-50 C	&	30	&	1460	&	947	&	30	&	1750	&	1330	&	\nodata	&	sat	&	sat	&	\nodata	&	sat			&	sat  		\\
K3-50 D	&	36	&	2240	&	1850	&	36	&	2580	&	2140	&	36	&	12600	&	9100	&	\nodata	&	sat			&	sat  		\\
K3-50 G	&	24	&	468		&	176 	&	24	&	421	    &	149  	&	24	&	3860	&	1620	&	24	&	8040		&	3710  	\\
\enddata
\tablecomments{If there is no $F_{\rm int-bg}$ value for a source, then the source is not well resolved from other nearby sources and/or extended emission. For these sources, the $F_{\rm int}$ value is used as the upper limit in the SED modeling. Entries with ``sat'' means they are saturated in that band. We use the point source saturation fluxes of 190, 200, 1400, and 740\,mJy at 3.6, 4.5, 5.8, and 8.0\,$\mu$m, respectively (from the Spitzer Observer’s Manual, Version 7.1.), as lower limits in the SED modeling.}
\tablenotetext{a}{Most of the field containing K3-50 is contaminated with array artifacts at 5.8 and 8\,$\mu$m due to the saturated pixels from Source A adding extra uncertainty to the 5.8 and 8\,$\mu$m photometry of all sources.}
\label{tb:IRAC_K350}
\end{deluxetable*}

\begin{deluxetable*}{lrrrrrr}
\tabletypesize{\scriptsize}
\tablecolumns{7}
\tablewidth{0pt}
\tablecaption{Herschel-PACS Observational Parameters of Sources in DR7}
\tablehead{\colhead{  }&
           \multicolumn{3}{c}{${\rm 70\mu{m}}$}&
           \multicolumn{3}{c}{${\rm 160\mu{m}}$}\\
           \cmidrule(lr){2-4} \cmidrule(lr){5-7}\\
           \colhead{ Source }&
           \colhead{ $R_{\rm int}$ } &
           \colhead{ $F_{\rm int}$ } &
           \colhead{ $F_{\rm int-bg}$ } &
           \colhead{ $R_{\rm int}$ } &
           \colhead{ $F_{\rm int}$ } &
           \colhead{ $F_{\rm int-bg}$ } \\
	   \colhead{  } &
	   \colhead{ ($\arcsec$) } &
	   \colhead{ (Jy) } &
	  \colhead{ (Jy) } &
	   \colhead{ ($\arcsec$) } &
	   \colhead{ (Jy) } &
	   \colhead{ (Jy) } \\
}
\startdata
\multicolumn{7}{c}{Compact Sources} \\
\hline
DR7 2	&16.0	&25.4	      & \nodata	    &22.5	&168	    &\nodata      \\
DR7 4	&16.0	&180	          & 38.2	    &22.5	&202	    &\nodata  \\
DR7 5	&16.0	&151	          & \nodata	    &22.5	&244    	&\nodata \\
DR7 6	&16.0	&187      	  & \nodata	    &22.5	&201       &\nodata  \\
\hline
\multicolumn{7}{c}{Extended Sources} \\
\hline
DR7 1	&16.0	&124	          & 31.3	    &22.5	&168    	&\nodata \\
DR7 3	&32.0	&750	          & 139    	&32.0	&500    	&146 \\
DR7 A	&22.5	&394	          & 107    	    &22.5	&196	    & 47.5 \\
DR7 B	&16.0	&266	          & 182	    &22.5	&236    	&111 \\
DR7 C	&28.8	&694              & 351    	    &35.2	&673	    &363 \\
DR7 F	&19.2	&187	          & 101    	&22.5	&168    	&\nodata  
\enddata
\tablecomments{If there is no $F_{\rm int-bg}$ value for a source, then the source is not well resolved from other nearby sources and/or extended emission. For these sources, the $F_{\rm int}$ value is used as the upper limit in the SED modeling.}
\label{tb:PACS_DR7}
\end{deluxetable*}

\begin{deluxetable*}{lrrrrrr}
\tabletypesize{\scriptsize}
\tablecolumns{7}
\tablewidth{0pt}
\tablecaption{Herschel-PACS Observational Parameters of Sources in K3-50}
\tablehead{\colhead{  }&
           \multicolumn{3}{c}{${\rm 70\mu{m}}$}&
           \multicolumn{3}{c}{${\rm 160\mu{m}}$}\\
           \cmidrule(lr){2-4} \cmidrule(lr){5-7}\\
           \colhead{ Source }&
           \colhead{ $R_{\rm int}$ } &
           \colhead{ $F_{\rm int}$ } &
           \colhead{ $F_{\rm int-bg}$ } &
           \colhead{ $R_{\rm int}$ } &
           \colhead{ $F_{\rm int}$ } &
           \colhead{ $F_{\rm int-bg}$ } \\
	   \colhead{  } &
	   \colhead{ ($\arcsec$) } &
	   \colhead{ (Jy) } &
	  \colhead{ (Jy) } &
	   \colhead{ ($\arcsec$) } &
	   \colhead{ (Jy) } &
	   \colhead{ (Jy) } \\
}
\startdata
\multicolumn{7}{c}{Compact Sources} \\
\hline
K3-50 1	&	16.0	&	143	&	\nodata	&	22.5	&	317	&	\nodata	\\
K3-50 2	&	16.0	&	228	&	\nodata	&	22.5	&	479	&	\nodata	\\
K3-50 3	&	16.0	&	262 	&	\nodata	&	22.5	&	608	&	\nodata	\\
K3-50 4	&	16.0	&	247	&	\nodata	&	22.5	&	398	&	\nodata	\\
K3-50 5	&	16.0	&	948	&	\nodata	&	22.5	&	1690	&	\nodata	\\
K3-50 6	&	16.0	&	372  	&	85.0	&	22.5	&	504	&	\nodata	\\
K3-50 7	&	16.0	&	1000	&	\nodata	&	22.5	&	2110	&	\nodata	\\
K3-50 8	&	16.0	&	340	&	\nodata	&	22.5	&	466	&	\nodata	\\
K3-50 C1	&	22.5	&	1320	&	1170	&	22.5	&	1240	&	1120	\\
K3-50 C2	&	22.5	&	744	   &	664  	&	22.5	&	599 	&	507 	\\
\hline
\multicolumn{7}{c}{Extended Sources} \\
\hline
K3-50 A	&	32.0	&	9120	&	8660	&	32.0	&	5150	&	4990	\\
K3-50 B	&	38.4	&	2360	&	1750	&	38.4	&	1540	&	1140	\\
K3-50 C	&	48.0	&	2690	&	2470	&	48.0	&	2530 	&	2130 	\\
K3-50 D	&	38.4	&	1690	&	1430	&	38.4	&	749 	&	516 	\\
K3-50 G	&	25.6	&	328 	&	231 	&	25.6	&	292 	&	187 	\\
\enddata
\tablecomments{If there is no $F_{\rm int-bg}$ value for a source, then the source is not well resolved from other nearby sources and/or extended emission. For these sources, the $F_{\rm int}$ value is used as the upper limit in the SED modeling.}
\label{tb:PACS_K350}
\end{deluxetable*}

\end{document}